\documentclass[journal]{new-aiaa}
\usepackage[utf8]{inputenc}

\usepackage{graphicx}
\usepackage{amsmath}
\usepackage{dsfont}
\usepackage[version=4]{mhchem}
\usepackage{siunitx}
\usepackage{longtable,tabularx}
\newcolumntype{R}{>{\raggedleft\arraybackslash}X}
\usepackage{tikz}
\usepackage{algorithm}
\usepackage{algpseudocode}
\usepackage{booktabs}
\usepackage{xtab}
\usepackage{graphicx}
\usepackage{subcaption}
\usepackage[normalem]{ulem}

\usetikzlibrary{shapes.geometric, arrows}
\tikzstyle{startstop} = [rectangle, rounded corners, minimum width=3cm, minimum height=1cm,text centered, draw=black, fill=gray!10]
\tikzstyle{process} = [rectangle, minimum width=3cm, minimum height=1cm, text centered, draw=black, fill=gray!10]
\tikzstyle{decision} = [diamond, minimum width=3cm, minimum height=1cm, text centered, draw=black, fill=gray!20]
\tikzstyle{arrow} = [thick,->,>=stealth, rounded corners]
\title{Uncertainty Quantification Study of a Re-entry Breakup}

\author{Tommy Williamson\footnote{Corresponding Author, PhD Student, Aerospace Centre of Excellence, University of Strathclyde, tommy.williamson@strath.ac.uk,  Student Member AIAA}}
\affil{University of Strathclyde, Glasgow, Scotland, G1 1XJ}
\author{Beatriz Jilete\footnote{Space Debris Monitoring Systems Engineer, Space Debris Office, GMV at European Space Agency}}
\affil{GMV at ESA ESAC, Villanueva de la Cañada, Spain, 28692}
\author{Emma Stevenson\footnote{Space Debris Engineer, Space Debris Office, IMS Space Consultancy at European Space Agency}}
\affil{IMS Space Consultancy at ESA ESTEC, Netherlands, 2200 AG}
\author{Stijn Lemmens\footnote{Senior Space Debris Mitigation Analyst, Space Debris Office, European Space Agency}}
\affil{European Space Agency, ESTEC, Netherlands, 2200 AG}
\author{Massimiliano Vasile\footnote{Professor, Aerospace Centre of Excellence, University of Strathclyde, Senior Member AIAA}, Marco Fossati\footnote{Professor, Aerospace Centre of Excellence, University of Strathclyde, Member AIAA}}
\affil{University of Strathclyde, Glasgow, Scotland, G1 1XJ}

\begin{document}
\onecolumn
\maketitle

\begin{abstract}
The uncertainty associated with breakup events that occur during atmospheric re-entry is severe. Limited attempts to gain a better knowledge of this environment have included the use of breakup recorder-type sensor capsules that are designed to escape the demising debris cloud and survive in order to transmit data. This work models a breakup recorder undergoing this process as a rigid body experiencing hypersonic aerothermodynamic loads alongside collision dynamics with components of the demising container vehicle. The re-entry of the Edoardo Amaldi Automated Transfer Vehicle (ATV3) and the recorder placed on board, the Re-Entry Breakup Recorder 4 (REBR4) is studied in the present work. After a deterministic exploration of the nature of the dynamics of the problem, uncertainty quantification is performed to investigate the effects of initial spacecraft state, REBR detachment conditions and spacecraft fragmentation states. From this data, inferences about the nature of the real re-entry event indicate that detachment of the recorder from the cargo bay prior to main breakup events is more likely than the alternate hypothesis of the container vehicle experiencing high rotation rates. 
\end{abstract}

\section*{Nomenclature}

{\renewcommand\arraystretch{1.0}
\noindent\begin{xtabular*}{\columnwidth}{@{}l @{\quad=\quad} l@{}}
$\beta$ & stabilisation tuning parameter \\
$\vec{\mathbf{d}}$ & intersection depths vector \\
$\Delta t$ & time step \\
$C_p$ & pressure coefficient \\
$C_\tau$ & shear coefficient\\
$e$ & coefficient of restitution \\
$J$ & constraint Jacobian \\
$\text{Kn}$ & Knudsen Number \\
$l$ & lower bound \\
$\ell_\text{ref}$ & reference length \\
$\vec{\mathbf{\lambda}}$ & impulse vector \\
$m$ & number of collision contacts \\
$M^{-1}$ & inverse mass matrix \\
$\mu$ & mean \\
$n$ & number of colliding bodies \\
$\hat{\mathbf{n}}$ & local normal \\
$\vec{\mathbf{\omega}}$ & rotational velocity vector \\
$\sigma$ & standard deviation \\
$\vec{\mathbf{v}}$ & translational velocity vector \\
\end{xtabular*}}
\section{Introduction}
\lettrine{A}tmospheric destructive re-entry is an extremely complex event that couples different physical processes and bears significant uncertainties\cite{wu_space_2011}. Spacecraft breakup during re-entry, alongside the dynamics and aerothermodynamics of the resulting cloud of debris, are key drivers of on-ground risk due to re-entry. When considering compliance verification procedures of spacecraft design, robust estimation of the casualty risk presented by uncontrolled re-entry of a satellite is vital to current and future sustainable use of space. Especially as the rate of satellite re-entry continues to accelerate yearly\cite{space_env_report, pardini_uncontrolled_2019}. The ability to appropriately design and plan the demise of a satellite according to the Design-For-Demise (D4D) guidelines \cite{noauthor_esa_nodate_D4D, cattani_overview_2021} can significantly reduce the casualty risk and it has now become a cornerstone of any new spacecraft and satellite mission. As such, the ability to make robust and reliable predictions of breakup events and the associated dynamics of the debris clouds is an area of great research interest and importance \cite{de_persis_overview_2024,sanson_breakup_2019,park_separation_2020}. 

Significant efforts have been put forward to reduce and to quantify the uncertainty on the complex and multi-physical event that is destructive atmospheric re-entry. This includes efforts directed to advance modelling capabilities in key areas such as material and structural response \cite{turchi_thermochemical_2017,park_re-entry_2021}, nonequilibrium aerothermodynamics \cite{LEMAOUT2025126999,fossatigraham2025,Morgado2024} and orbital state/dynamics \cite{trisolini_propagation_2021,mehta_sensitivity_2017}. Improvements in the multi-physics modelling of destructive re-entry cannot alone fully resolve the problem of making reliable predictions. This is due to the high demand of computational resources that accompanies high-fidelity methods on one side, but also results from the inherent epistemic uncertainty of the physical and numerical models and the aleatoric uncertainty of the re-entry events themselves. Substantial work is therefore still needed, especially in relation to the ability of introducing a consistent treatment uncertainty quantification as a fundamental complement to physical modelling of destructive re-entry and demise events\cite{parigini2015debris, mehta_sensitivity_2017, pardini_assessing_2018, geul2018analysis}. From a flight dynamics perspective, the uncertainties in initial spacecraft state, breakup events, fragments collision and atmospheric density are recognised as key for quantifying the casualty risk associated with re-entry\cite{wu_space_2011,park_re-entry_2021}. The industry standard approach, as recommended by the European Space Agency's Space Debris Mitigation guidelines\cite{noauthor_esa_nodate}, uses Monte Carlo (MC) campaigns to ensure that the prediction of the demise of a satellite is robust in the presence of such uncertainties, such MC campaigns can be augmented using methods from statistics such as bootstrapping\cite{LEMMENS20152592}.

This work seeks to explore computationally the complex proximal dynamics resulting by explicitly accounting for the collision dynamics of rigid bodies subject to hypersonic aerothermal loads and by using uncertainty treatment on key parameters so that the impact of key modelling aspects on the resulting outcome of the re-entry event can be assessed. Focusing on these aspects is deemed to allow for a better insight into and interpretation of the extant or planned re-entry data collection efforts\cite{schmidt_kentucky_2023,jilete_draco_2025,feistel_comparison_2013} that recover information from a \textit{``data recorder''}  capsule designed to survive re-entry that transmits data through a satellite communications network. Being initially attached to a demising container vehicle, the recorder might separate at some point during breakup and it becomes important to be able to complement recorded data with simulations that allow assessing these separation events to avoid relating the dynamics of the recorder to the dynamics of some key component of the spacecraft, e.g. cargo or avionics bay or tanks. In this work comparison of the recorded data to simulations that model the complex dynamics of colliding rigid bodies in hypersonic flow enables inferences to be made on when during re-entry detachment and escape events occur.

Re-entry simulators are commonly classified into lower computation cost \textit{component-oriented} tools, such as ESA's DRAMA\cite{braun_drama_2020} or NASA's ORSAT\cite{greene_recent_2023} and higher computation cost \textit{spacecraft-oriented} tools such as ESA's SCARAB\cite{karrang_review_2025} and CNES's PAMPERO\cite{van_hauwaert_pampero_2022}. The key distinction here is that component-oriented codes are predicated on the assumption that a breakup process will disperse components of the vehicle in the atmosphere such that they can be modelled individually and independently from one another, whereas spacecraft-oriented codes model fragmentation as consequent to the state of the spacecraft. The \textbf{T}ransatmospher\textbf{I}c fligh\textbf{T} simulAtio\textbf{N} tool (TITAN)\cite{morgado_multifidelity_2024} is a recently developed open-source re-entry simulator implementing a ``multifidelity'' paradigm\cite{de_persis_overview_2024} capable of modelling re-entry on a sliding scale of model fidelities. Of particular note in this context is that TITAN can model interactions between free-flying fragments/components in a way that is unprecedented for re-entry simulation. Through the application of raytracing-based flow shadowing and a collision model, fragments can be affected dynamically and thermally by others.
\par 
The specific case analysed here is the re-entry of the Edoardo Amaldi Automated Transfer Vehicle 3 (ATV3)\cite{kinnersley_automated_2011}, equipped with the Re-Entry Breakup  Recorder 4 (REBR4)\cite{feistel_comparison_2013, ailor_reentry_2013}. Alongside being a popular case for exploration of re-entry modelling technologies, many works modelling re-entry of the ATV exist\cite{boutamine_computational_2007,bastida_virgili_simulation_2010,koppenwallner_analysis_2005, morgado_multifidelity_2024, graham_comparative_2026} that constitute the source of important reference data. However specific interest is often paid to rebuilding the ATV1 Jules Verne re-entry observation campaign\cite{de_pasquale_atv_2009,snively_airborne_2011}, as opposed to the ATV3. After undocking from the ISS on 21:44 GMT on 28 September 2012, the ATV3 re-entered the Earth's atmosphere on 3 October 2012. Unlike the previous REBR-ATV mission, the REBR4 successfully transmitted 386s of recorded data\cite{ailor_reentry_2013}, providing information on the temperatures, pressures and angular velocities experienced by the REBR that could be considered to be under-exploited by the re-entry modelling community.\par
\begin{figure}[t]
    \centering
    \includegraphics[width=\linewidth]{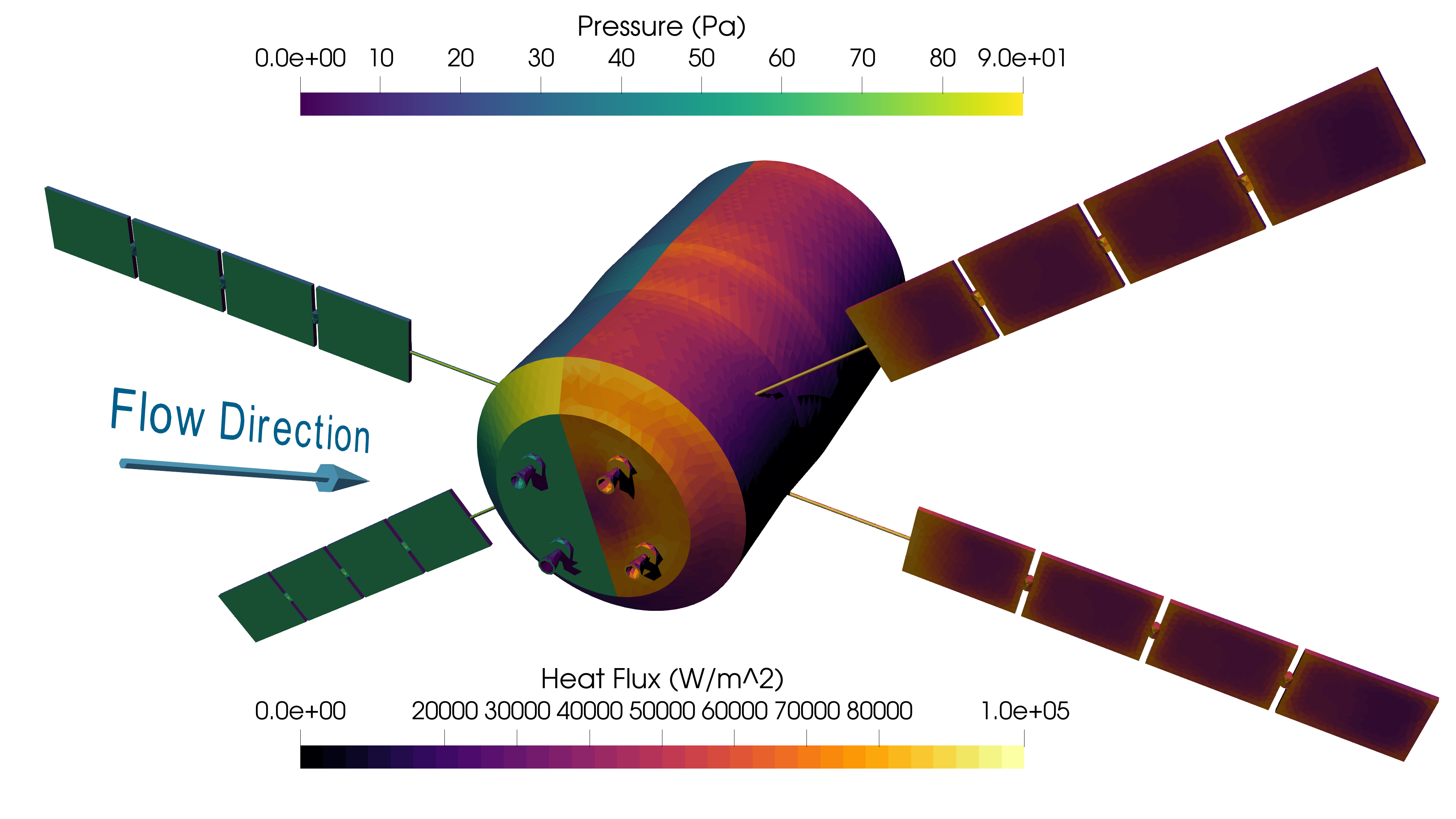}
    \caption{Surface pressure (left) and heat flux distributions (right) of the ATV prior to panel fragmentation}\label{fig:atv_mnt}
\end{figure}
This work is structured as follows. In \autoref{sec:method} an explication will be provided of the modelling technologies specifically within TITAN applied to this case. Following this, \autoref{sec:deter} details the setup and results of a deterministic exploration of the problem. After which in \autoref{sec:uq} a Monte Carlo campaign will be described that analyses the case from a stochastic perspective. Finally, conclusions will be presented in \autoref{sec:conc}.

\section{Simulating Re-entry Multiphysics with TITAN}\label{sec:method}
TITAN is an open-source multifidelity multiphysics simulator designed with the core focus and capability of simulating destructive re-entry. In order to  represent the behaviour of a demising spacecraft as completely as possible, TITAN computes the following phenomena: aerodynamic pressures and loads; aerothermodynamic incident heat flux; thermal response and ablation; 6-Degree-of-Freedom (6DoF) flight dynamics and rigid body collision dynamics. These phenomena and the methodologies/codes used to resolve them can be seen in \autoref{tab:TITAN_overview}, with the methods deployed in this analysis highlighted in \textbf{bold}. Note that for computational efficiency low-fidelity methods are preferred unless considered specifically relevant to the problem. Details of the relevant models are presented below.
\begin{table*}[b]
  \footnotesize
  \caption{TITAN Multifidelity Simulation Options}
  \label{tab:TITAN_overview}
  \centering
  \begin{tabular*}{\linewidth}{@{\extracolsep{\fill}}lll}
    \toprule
    \textbf{Discipline} & \textbf{Low Fidelity} & \textbf{High Fidelity}\\
    \midrule
    Aerodynamics & \textbf{Schaaf-Chambre/Modified Newtonian Theory Panel Codes} & SU2-NEMO CFD\cite{maier_su2-nemo_2021}/SPARTA DSMC\cite{plimpton_direct_2019}\\
    Aerothermodynamics & \textbf{Heat flux relations} & SU2-NEMO CFD\cite{maier_su2-nemo_2021}/SPARTA DSMC\cite{plimpton_direct_2019} \\
    Thermal Response & \textbf{Lumped Mass Model} & PATO Material Response Code\cite{lachaud_porous-material_2014} \\
    6DoF Dynamics & Multi-step Integrators & \textbf{Adaptive RK Integrators} \\
    Collision Resolution & Baumgarte Stabilisation & \textbf{Split-Impulse Stabilisation} \\
    \bottomrule
  \end{tabular*}
  \end{table*}

\subsection{Resolving Hypersonic Aerothermodynamics}
The core of TITAN's low-fidelity aerodynamic model is a raytracer that recovers surface pressure and shear distributions according to an appropriate panel code for the flow regime at hand, i.e. Schaaf-Chambre\cite{schaaf_flow_1958} for free-molecular flow and Modified Newtonian Theory (MNT)\cite{lees_hypersonic_2003} for continuum flow, an example of the pressure and heat flux distributions produced by the aerodynamic model can be seen in \autoref{fig:atv_mnt}. The raytracer computes an angle of incidence ($\theta$) between the freestream flow direction ($\hat{\mathbf{v}}_{\infty}$) and local normal $\hat{\mathbf{n}}$ of each panel that \textit{``sees''} the flow.
\begin{equation}
    \theta = \cos^{-1}({\hat{\mathbf{v}}_{\infty} \cdot \hat{\mathbf{n}}})
\end{equation}
The pressure and shear distributions are then found as solely functions of freestream conditions and $\theta$. For example in the MNT case.
\begin{equation}
    C_p^\text{MNT} = 
    \begin{cases}
        {C_p}_\text{max} \sin ^2 (\theta) \quad&\forall \theta>0\\
        0 \quad&\text{otherwise}
    \end{cases}
      \quad C_\tau^\text{MNT} = 0
\end{equation}
Where ${C_p}_\text{max}$ is the stagnation pressure coefficient. A Knudsen-based bridging function averages between Schaaf-Chambre and MNT in the transitional regime\cite{mehta_open_2015}. A family of sigmoid curves are constructed from data for a range of different effective nose radii, interpolation is then used to acquire a function that smoothly transitions between free-molecular and continuum models for the effective nose radius of the vehicle. Heat fluxes can then be derived from the resultant distributions\cite{schaaf_flow_1958,fay_theory_1958}. It should be noted that MNT is an inviscid model that neglects any shear or skin friction terms and that therefore in the continuum regime rotational motion in TITAN is undamped, aside from any damping resulting from shape effects on the pressure field.

The large principal dimensions of the ATV means that even at the nominal entry interface selected here of $129$km the container vehicle experiences Knudsen numbers on the order of $\text{Kn}=10^{0}$, well below the defined free-molecular limit in TITAN of $10^2$, with the vast majority of the entry profile occurring below TITAN's defined continuum limit at $\text{Kn}\leq 10^{-3}$. 
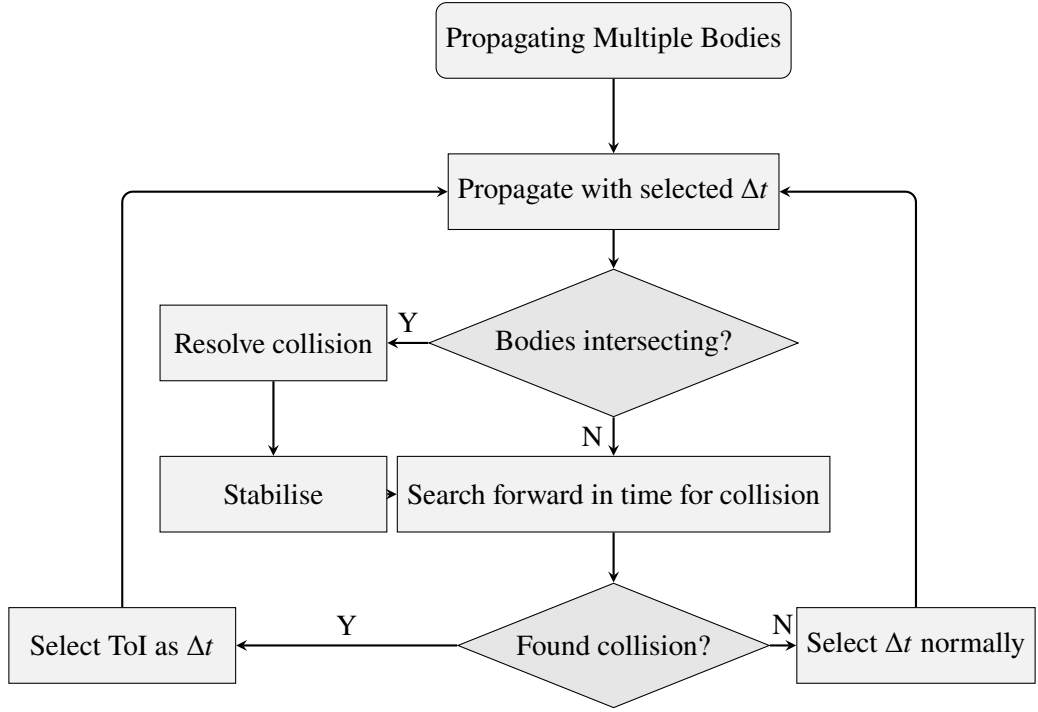
\begin{figure}[h]
    \centering
    \scalebox{1.0}{%
        \begin{tikzpicture}[node distance=2cm]
            \node (start) [startstop] {Propagating Multiple Bodies};
            \node (prop) [process, below of=start] {Propagate with selected $\Delta t$};
            \node (coll) [decision, aspect=2.5, below of=prop] {Bodies intersecting?};
            \node (toi) [process, below of=coll] {Search forward in time for collision};
            \node (toicoll) [decision, aspect=2.5, below of=toi] {Found collision?};
            \node (toiasdt) [process, left of=toicoll,  xshift=-4.5cm] {Select ToI as $\Delta t$};
            \node (normaldt) [process, right of=toicoll,  xshift=2.0cm] {Select $\Delta t$ normally};
            \node (resolve) [process, left of=coll, xshift=-2.5cm] {Resolve collision};
            \node (stab) [process, below of=resolve] {Stabilise};
            \draw [arrow] (start) -- (prop);
            \draw [arrow] (prop) -- (coll);
            \draw [arrow] (coll) -- node[anchor=east]{N} (toi);
            \draw [arrow] (coll) -- node[anchor=south]{Y} (resolve);
            \draw [arrow] (resolve) -- (stab); 
            \draw [arrow] (stab) -- (toi); 
            \draw [arrow] (toi) -- (toicoll);
            \draw [arrow] (toicoll) --node[anchor=south]{Y} (toiasdt);
            \draw [arrow] (toicoll) --node[anchor=south]{N} (normaldt);
            \draw [arrow] (toiasdt) |- (prop);
            \draw [arrow] (normaldt) |- (prop);
        \end{tikzpicture}}
    \caption{Overview of the collision model implemented in TITAN}\label{fig:coll_flowchart}
\end{figure}
\subsection{Collision Modelling}
 The nature of a dispersive breakup event is dependent on the geometric and kinematic interaction of components within the vehicle and debris cloud, which TITAN resolves using a multibody kinematics solver. Computational rigid body dynamics accounting for collision and contact accurately model kinematic reactions of objects in cases of infinitesimal or negligible deformation. As such they are often deployed in contexts such as computer graphics, video game development and especially in robotics research\cite{park_geometric_2018}. In contrast to elastic or smooth methods\cite{flores_contact_2022}, rigid methods are also called non-smooth in that discontinuities are introduced in order to resolve contacts. These discontinuities are present either in position\cite{deul_positionbased_2016} or in velocity through applied impulses. Whilst position-based dynamics are in certain cases desirable for stability and efficiency, impulsive methods are often preferred in contexts without real-time restrictions. Popular impulsive physics engines such as MuJoCo\cite{todorov_mujoco_2012}, the Open Dynamics Engine (ODE)\cite{brugali_extending_2014} and Bullet\cite{coumans_bullet_2015} use a \textit{constraint-solving} approach that solves for the necessary impulses to produce a solution that lies on the constraint manifold of the problem. The specific program employed was introduced by Lötstedt\cite{lotstedt_coulomb_1981,lotstedt_numerical_1984} following the work of Moreau\cite{moreau_quadratic_1966}, formulating this problem as a \textit{linear complementarity problem} (LCP)\cite{anitescu_formulating_1997}, as described by Cottle and Danzig\cite{cottle_complementary_1968}.
The methodology used here is a Projected Gauss-Siedel (PGS) LCP collision solver with binary search time-of-impact (ToI) collision detection and Baumgarte\cite{baumgarte_stabilization_1972} or split-impulse stabilisation\cite{{ascher_stabilization_1995,brugali_extending_2014,coumans_bullet_2015}}. Implementations of collision models can vary across different physics engines\cite{coumans_bullet_2015,todorov_mujoco_2012,brugali_extending_2014,baraff_fast_1994,deul_positionbased_2016}, each with different strengths and weaknesses, especially in terms of the inescapable fidelity-efficiency tradeoff. As such, it was considered appropriate to provide a brief overview of the specific model used in the code (\autoref{fig:coll_flowchart}) in order to understand the fidelity level with which collisions are modelled in TITAN. Where bodies may collide with each other (i.e. are part of the same debris cloud) TITAN propagation uses Time-of-Impact prediction to detect collisions (\autoref{ssec:ToI}) before resolving them by solving the constrained dynamics problem of contacting rigid bodies (\autoref{ssec:Reso}). Further to this, constraint violations are remedied with stabilisation methods (\autoref{ssec:Stab}).

\subsubsection{Time-of-Impact Prediction}\label{ssec:ToI}
Simple collision detection algorithms, where collisions are checked at each time step, are plagued by the problem of \textit{tunnelling}\cite{stewart_implicit_1996}, where an object of small principal dimension or large velocity may ``skip'' an intersection that would be flagged if it occurred ``\textit{on}'' a time step since the actual transition of non-contact to contact occurs ``\textit{between}'' time steps. Such a problem is challenging because although it is obvious as a \textit{posteriori} observer that a collision has been missed it is not trivial to detect such collisions \textit{a priori}.

TITAN ameliorates this problem using a binary search method to identify the exact time of impact such that the next dynamical update will result in a minimal non-zero intersection, that can be resolved by impulsive collision resolution. The ToI is found through linear extrapolation of the relative motion of the debris cloud and mesh intersection checks. In actuality the motion of a hypersonic debris cloud cannot be accurately described as linear. As such, a limit must be placed on propagation time step such that, for the purposes of collision prediction, nonlinear effects can be considered negligible. This limit is defined as the time taken for the object of highest relative velocity to traverse the reference length of the smallest object.
\begin{equation}
    \Delta t_\text{max} = \frac{\text{min}[\ell_\text{ref}]}{\text{max}[\vec{\mathbf{v}}_\text{rel}]}
\end{equation}
For a typical re-entry debris cloud, this value is on the order of $\sim10^{-1}$ seconds whilst fragments remain proximal. Once a collision has been predicted and propagated to, TITAN resolves it.
\subsubsection{Impulsive Collision Resolution}\label{ssec:Reso}
For $n$ rigid bodies participating in collisions with $m$ contacts in 3D-space (with 6 degrees of freedom), motion can be described with generalised velocity vector $\vec{\mathbf{v}}\in \mathds{R}^{6n}$.
\begin{equation}
    \vec{\mathbf{v}} = (v^x_0,v^y_0,v^z_0, \omega^x_0, \omega^y_0, \omega^z_0, \dots, v^x_n,v^y_n,v^z_n, \omega^x_n, \omega^y_n, \omega^z_n)^T
\end{equation}
Where $\vec{\mathbf{v}}_i$ and $\vec{\mathbf{\omega}}_i$ are the translational and rotational velocity vectors of the $i$th body in the Earth-centered Earth-fixed (ECEF) frame respectively. This generalised velocity vector can be related to the necessary impulse vector $\vec{\mathbf{\lambda}}\in \mathds{R}^{6n}$ to maintain constrained motion with the inverse mass matrix $M^{-1}\in \mathds{R}^{6n \times 6n}$, constraint Jacobian $J \in \mathds{R}^{6n \times m}$, contact normal velocity vector $\vec{\mathbf{v}}_N\in \mathds{R}^{m}$ and coefficient of restitution $e$.

\begin{equation}
    \vec{\mathbf{\lambda}} = {[JM^{-1}J^T]}^{-1}  (\vec{\mathbf{v}}_N - eJ\vec{\mathbf{v}})
\end{equation}
This problem can be formatted as an LCP. For clarity, let $\vec{\mathbf{v}}_N =\vec{\mathbf{y}}$, $JM^{-1}J^T = A$, $\vec{\mathbf{\lambda}} = \vec{\mathbf{x}}$ and $eJ\vec{\mathbf{v}} =\vec{\mathbf{b}}$.
\begin{equation}
    \begin{split}
    \vec{\mathbf{v}}_N = JM^{-1}J^T \vec{\mathbf{\lambda}}+eJ\vec{\mathbf{v}} \quad \equiv \quad \vec{\mathbf{y}} = \mathbf{A}\vec{\mathbf{x}} +&\vec{\mathbf{b}} \\ 
    \text{s.t.}\quad \mathbf{A}\vec{\mathbf{x}}\geq 0 \perp 0 \leq &\vec{\mathbf{x}}\quad
    \end{split}
\end{equation}
Where $\mathbf{A}\vec{\mathbf{x}}\geq 0 \perp 0 \leq \vec{\mathbf{x}}$ describes the \textit{complementarity constraint}, equivalently stated in plain English as follows: \textit{where $\mathbf{A}\vec{\mathbf{x}}$ is greater than zero, $\vec{\mathbf{x}}$ is zero and where  $\vec{\mathbf{x}}$ is greater than zero, $\mathbf{A}\vec{\mathbf{x}}$ is zero}. This constraint has an intuitive physical meaning in terms of contact force and separation, at all points where contact separation is zero the contact force is non-zero and all points where contact separation is non-zero the contact force is zero. The Projected Gauss-Siedel method solves the LCP according to a conventional Gauss-Siedel program with an additional ``projection'' step to ensure complementarity. It was found for the types of contact problem present in re-entry rigid body dynamics that PGS solutions for the contact impulses reliably converged in $\sim50$ iterations, presenting a tolerable minor computational overhead with respect to the more intensive parts of the TITAN simulation, e.g. the afore-mentioned ray-tracing for surface pressure determination. 

Once appropriate contact impulses have been determined, the necessary instantaneous velocity change, $\Delta \vec{\mathbf{v}}$, can be found.
\begin{equation}
    \Delta \vec{\mathbf{v}} = M^{-1}J^T \vec{\mathbf{\lambda}}
\end{equation}
By mapping $\vec{\mathbf{v}} + \Delta \vec{\mathbf{v}}$ back to the respective velocities of participating bodies the dynamical propagation can resume, with a new contact problem formulated and solved for each collision ensemble that occurs.
\subsubsection{Stabilisation}\label{ssec:Stab}
Whilst updates to bodies participating in collisions fully resolve the kinematics of the instantaneous contact that occurs, aerodynamic forces are not accounted for. As such no guarantee can be placed upon intersection-free dynamics due to impulsive collision modelling in TITAN, therefore stabilisation methods are made available to the user to increase the robustness of the model. 
The most straightforward and robust method is Baumgarte stabilisation, where a corrective velocity, $\Delta \vec{\mathbf{v}}_\text{corr}$, is applied in addition to the true physical response.
\begin{equation}
    \vec{\mathbf{v}}_{\text{post}} = \vec{\mathbf{v}}_{\text{prior}} + \Delta \vec{\mathbf{v}}_\text{phys} + \Delta \vec{\mathbf{v}}_\text{corr}
\end{equation}
$\Delta \vec{\mathbf{v}}_\text{corr}$ is computed from a vector of contact impulses, $\vec{\mathbf{\lambda}}_\text{corr}$, that scale with the vector of contact intersection depths, $ \vec{\mathbf{d}}$, according to a tuning parameter, $\beta \in [0,1]$. A value of $\sim0.2$ is often suggested to be optimally stable for $\beta$.
\begin{equation}
    \vec{\mathbf{\lambda}}_\text{corr} =[JM^{-1}J^T]^{-1}\cdot (-\beta\vec{\mathbf{d}})
\end{equation}
Of course such a relation can be solved using the same methods as applied for the \textit{``true''} response. It is intuitive to think of these stabilisation impulses as virtual springs of extensions $ \vec{\mathbf{d}}$ and stiffness $\beta / \Delta t$. Such a perspective reveals, simultaneously, the desirable feature of the method in terms of strongly recovering from intersections and the undesirable feature of adding unphysical energy to the system. It is for this reason that large values of $\beta$ can provoke significant instability.
Split-impulse stabilisation\cite{ascher_stabilization_1995,brugali_extending_2014,coumans_bullet_2015}, also known as \textit{``pseudo-velocities''} and \textit{``position projection''} is a modification of Baumgarte stabilisation that respects conservation. This is ensured by only carrying the physical velocity change forward in the simulation, the corrective term is applied and then discarded. The method as applied in TITAN differs slightly from conventional implementations, here the correction is applied as a small post-resolution \textit{``teleportation''} that instanteously adjusts body positions in-place.
\begin{equation}
    \begin{split}
    \vec{\mathbf{v}}_{\text{post}} &= \vec{\mathbf{v}}_{\text{prior}} + \Delta \vec{\mathbf{v}}_\text{phys}\\
    \vec{\mathbf{x}}_{\text{post}} &= \vec{\mathbf{x}}_{\text{prior}} + \Delta \vec{\mathbf{v}}_\text{corr}\Delta t
    \end{split}
\end{equation}
Where $\Delta \vec{\mathbf{v}}_\text{corr}$ is calculated as previously. The amelioration of conservation problems enables larger values of $\beta$ to be used safely, although at exceedingly large values oscillatory effects can still destabilise simulations. Robust performance in TITAN has been observed at $\beta\approx0.6$.
\section{Deterministic Simulation of ATV Re-entry}\label{sec:deter}
A \textit{nominal} set of initial conditions was selected to explore the problem (\autoref{ssec:Case}) in order to assess the nature of dynamics that potentially exist in this case and select relevant modelling paradigms to appropriately resolve the phenomena of interest. Considering the results of this nominal simulation a hypothesis was formed on the behaviour of the REBR (\autoref{ssec:Capsule}). 
\begin{figure*}[p]
    \centering
    \begin{subfigure}{0.45\linewidth}
         \centering
        \includegraphics[trim= 350 0 0 0, clip, width=\textwidth]{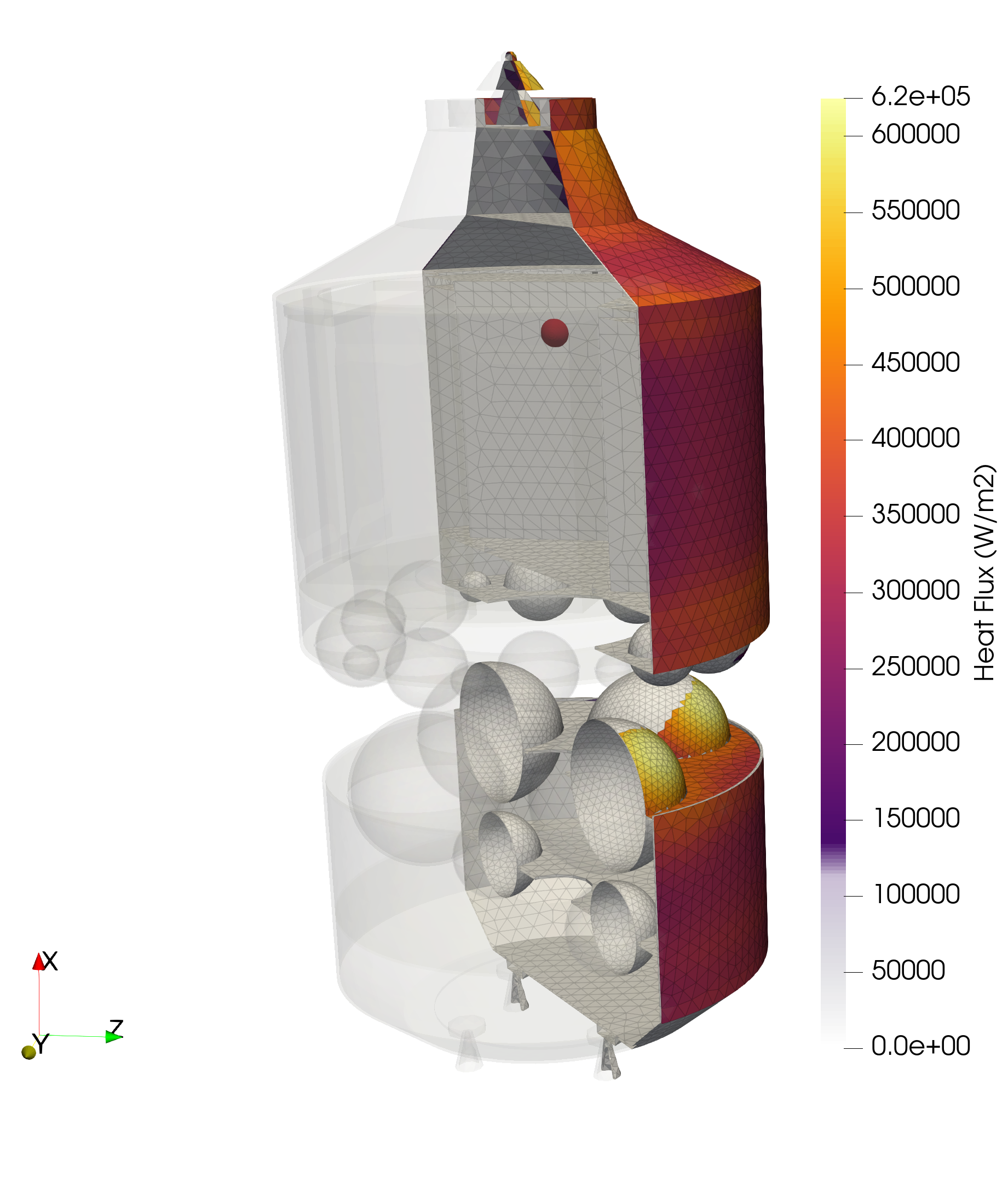}
        \caption{Propulsion Bay Separation}
        \label{subfig:sep_1}
    \end{subfigure}
    \begin{subfigure}{0.45\linewidth}
         \centering
        \includegraphics[trim= 350 0 0 0, clip, width=\textwidth]{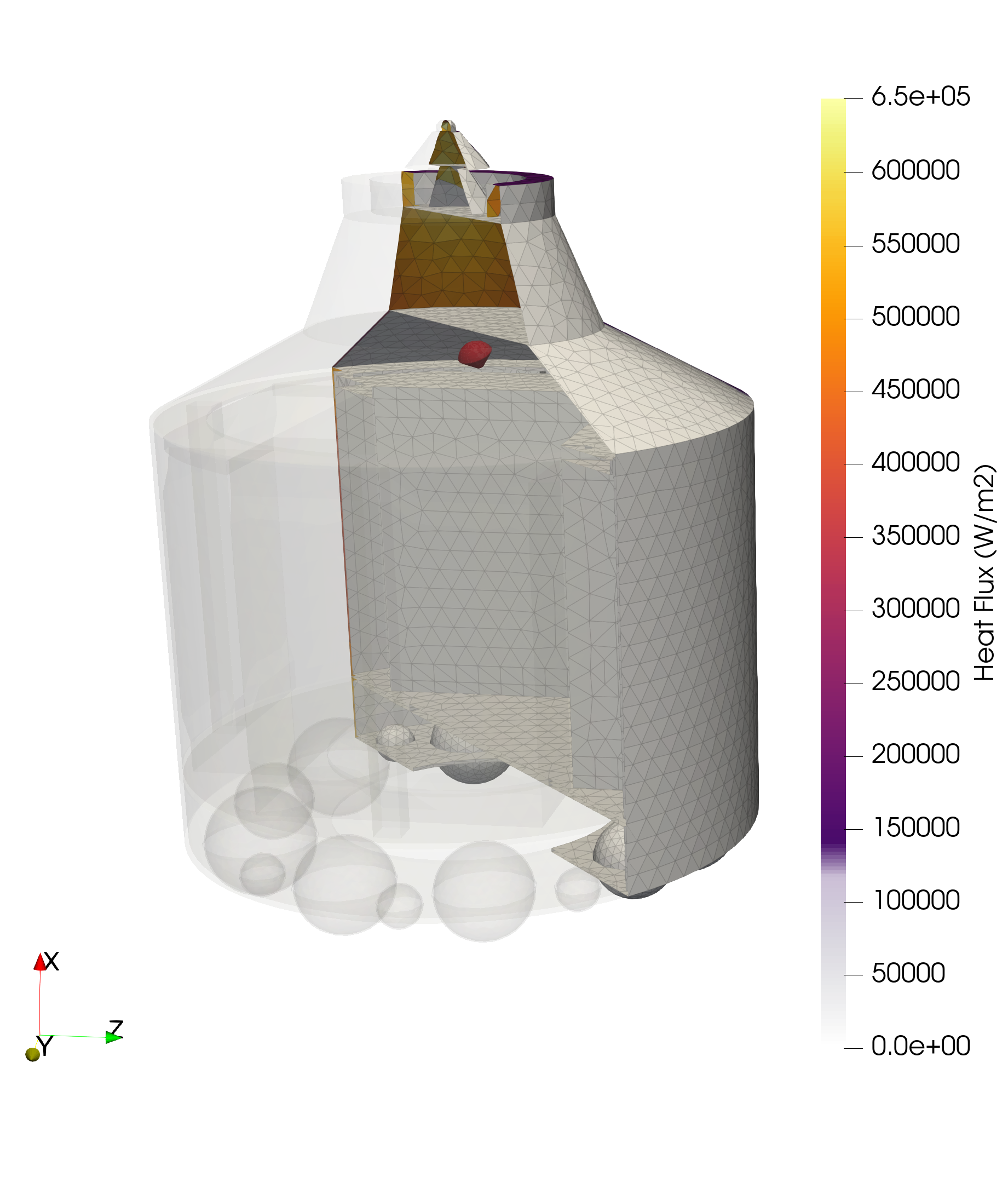}
        \caption{REBR Detachment}
        \label{subfig:sep_2}
    \end{subfigure}\\
    \begin{subfigure}{0.45\linewidth}
         \centering
        \includegraphics[trim= 350 0 0 0, clip, width=\textwidth]{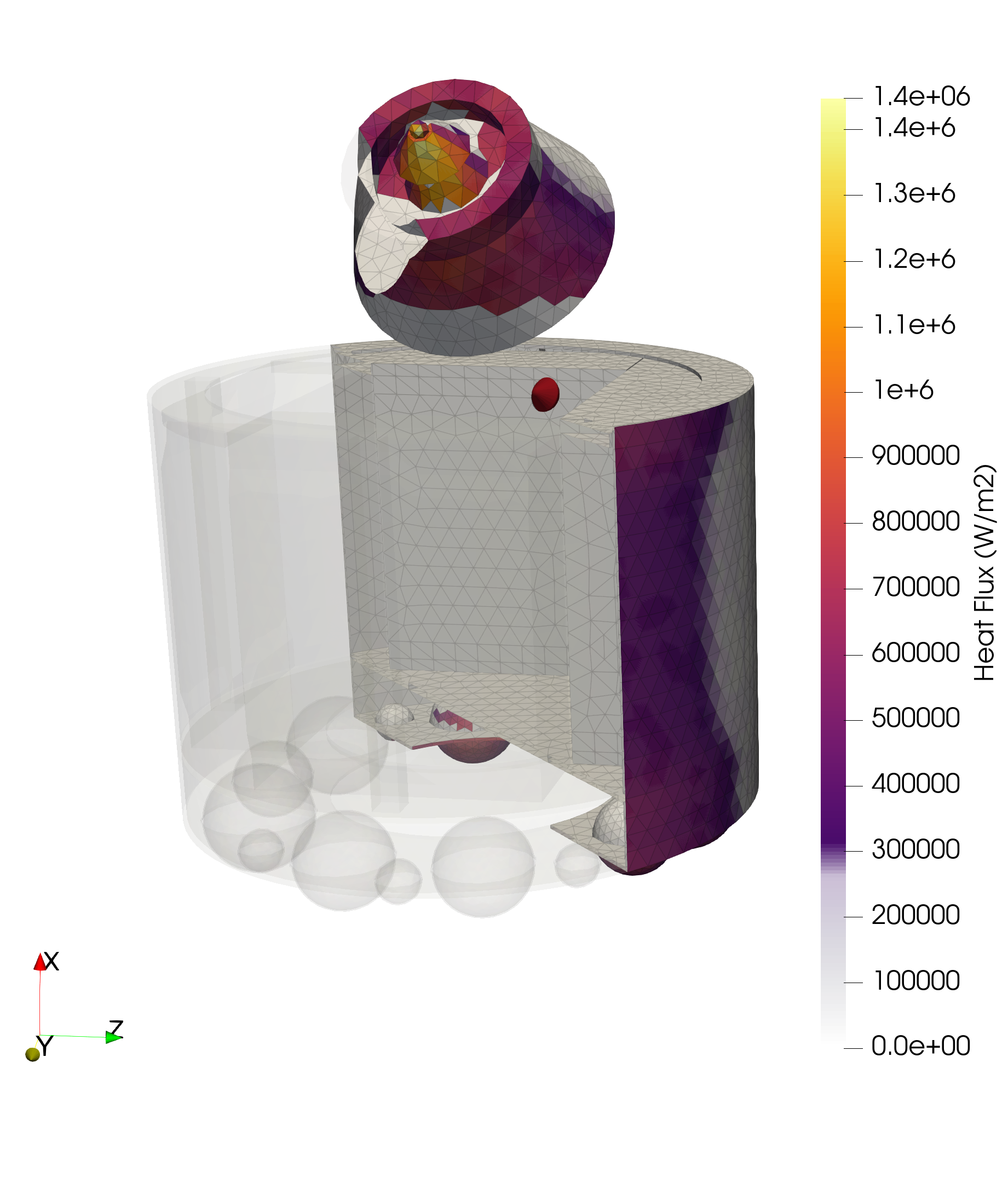}
        \caption{Equipped Bay Cap Separation}
        \label{subfig:sep_3}
    \end{subfigure}
    \caption{Nominal event instances in the equipped bay body frame (with REBR highlighted in red)}
    \label{fig:separations}
\end{figure*}
\subsection{Case Definition}\label{ssec:Case}
A mesh-based representation of the ATV was created. Of particular note is that the REBR was present as an individually simulated component, initially attached to the ATV prior to its \textit{``detachment''}. Afterwards it was modelled as free to move and collide inside the ATV's \textit{``equipped payload bay''} (EPB), represented in this work as a cuboidal-region bounded by equipment racks that were assumed to be entirely full. The mass properties of the intact ATV as simulated can be seen in \autoref{tab:ATV_mass} and a detailed breakdown of the representations of component materials is provided in \autoref{ap:materials}. The mass and material properties are derived from a generic ATV spacecraft configuration to be of sufficient accuracy for the dynamic analysis but do not necessarily correspond to the flown configuration. It is important to note that the material representation provided here cannot be definitive as the actual state of the ATV's equipment racks upon re-entry was not available for this research. Exploration of how different loading conditions might have affected re-entry is an interesting uncertainty quantification problem beyond the scope of this work.

The parameters defining the nominal case of the ATV at entry interface can be seen in \autoref{tab:ATV_case}, again it is important to note that the provided trajectory properties are derived from a generic ATV spacecraft configuration for a re-entry in the South Pacific Uninhabited Ocean Area to be of sufficient accuracy for the dynamic analysis but do not necessarily correspond to the flown trajectory. From this simulation starting point the container vehicle is simulated free-tumbling in transitional and continuum flow until fragmentation events occur. Pre-determined \textit{``triggers''} are selected \textit{a priori} to initiate demise or detachment of certain components based upon altitude (\autoref{tab:ATV_log}), these triggers were synthesized based upon matching the event log provided by Ailor\cite{ailor_reentry_2013} to the geometry and component fidelity available in the current TITAN representation of the ATV. The different stages of fragmentation can be seen in \autoref{fig:separations}.
\begin{table}[h]
    \caption{ATV3 Model Mass Properties}
    \label{tab:ATV_mass}
    \footnotesize
    \centering
    \begin{tabular}{llllll}
        \toprule
        & \textbf{Mass} (\unit{\kilogram}) & \textbf{CoG} (\unit{\metre}) & $\mathbf{I_{xx}}$ (\unit{\kilogram\metre\squared})& $\mathbf{I_{yy}}$(\unit{\kilogram\metre\squared})& $\mathbf{I_{zz}}(\unit{\kilogram\metre\squared})$\\
        \midrule
        With Solar Arrays & 16540 & 4.26 & 56130 & 104930 & 112160 \\
        Without Solar Arrays & 16325 & 4.29 & 44620 & 101950 & 100910 \\
        \bottomrule
  \end{tabular}
  \end{table}

\begin{table}[h]
  \caption{ATV3 Model Nominal Case Definition}
  \label{tab:ATV_case}
  \centering
  \begin{tabularx}{\columnwidth}{XR}
    \toprule
    \textbf{Parameter} & \textbf{Value}\\
    \midrule
    Altitude & 129.3638~\unit{\kilo\metre}\\
    Latitude & \ang{-21.1286}\\
    Longitude & \ang{-174.0972}\\
    Velocity & 7.5643~\unit{\kilo\metre\per\second}\\
    Flight Path Angle & \ang{-1.4678} \\
    Heading Angle & \ang{140.8619} \\
    Initial rotation & \ang{10} \unit{\per\second} (Pitch) \\
    \bottomrule
  \end{tabularx}
\end{table}
\begin{table}[h]
    \caption{ATV3 Nominal Event Log}
    \label{tab:ATV_log}
    \centering
    \begin{tabular}{lr}
        \toprule
        \textbf{Event} & \textbf{Trigger Altitude} \unit({km})\\
        \midrule
        Propulsion Bay separation & 73.0 (\autoref{subfig:sep_1})\\
        REBR Detachment & 71.2 (\autoref{subfig:sep_2})\\
        Equipped Bay Cap separation & 69.4 (\autoref{subfig:sep_3})\\

        \bottomrule
    \end{tabular}
\end{table}
\begin{figure}
    \centering
    \includegraphics[width=0.8\linewidth]{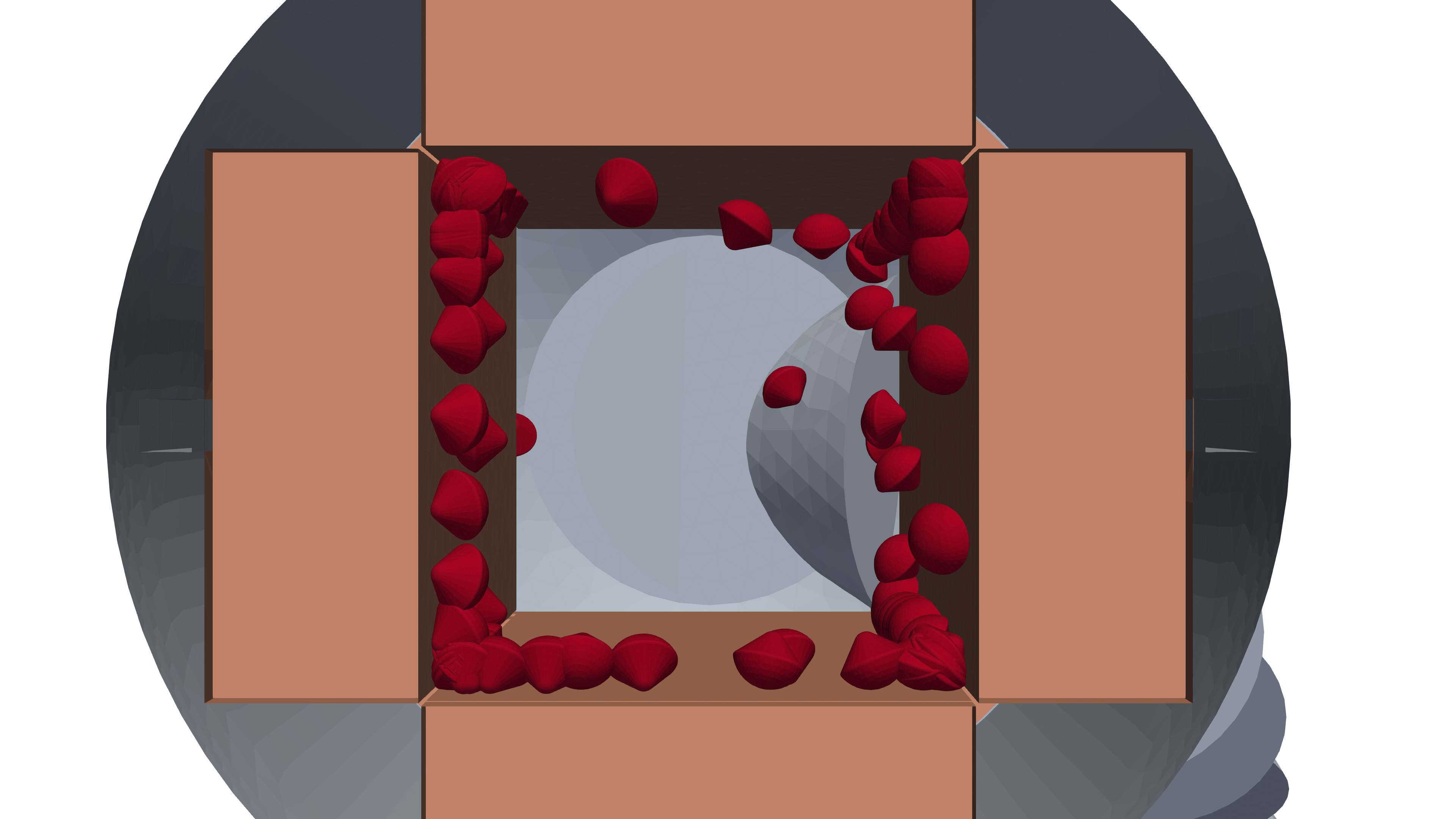}
    \caption{The motion of the REBR inside the payload bay ($\mathbf{\Delta t=0.32s}$, Bay frame)}
    \label{fig:rebrtime}
\end{figure}
\begin{figure}
    \centering
    \includegraphics[width=0.8\linewidth]{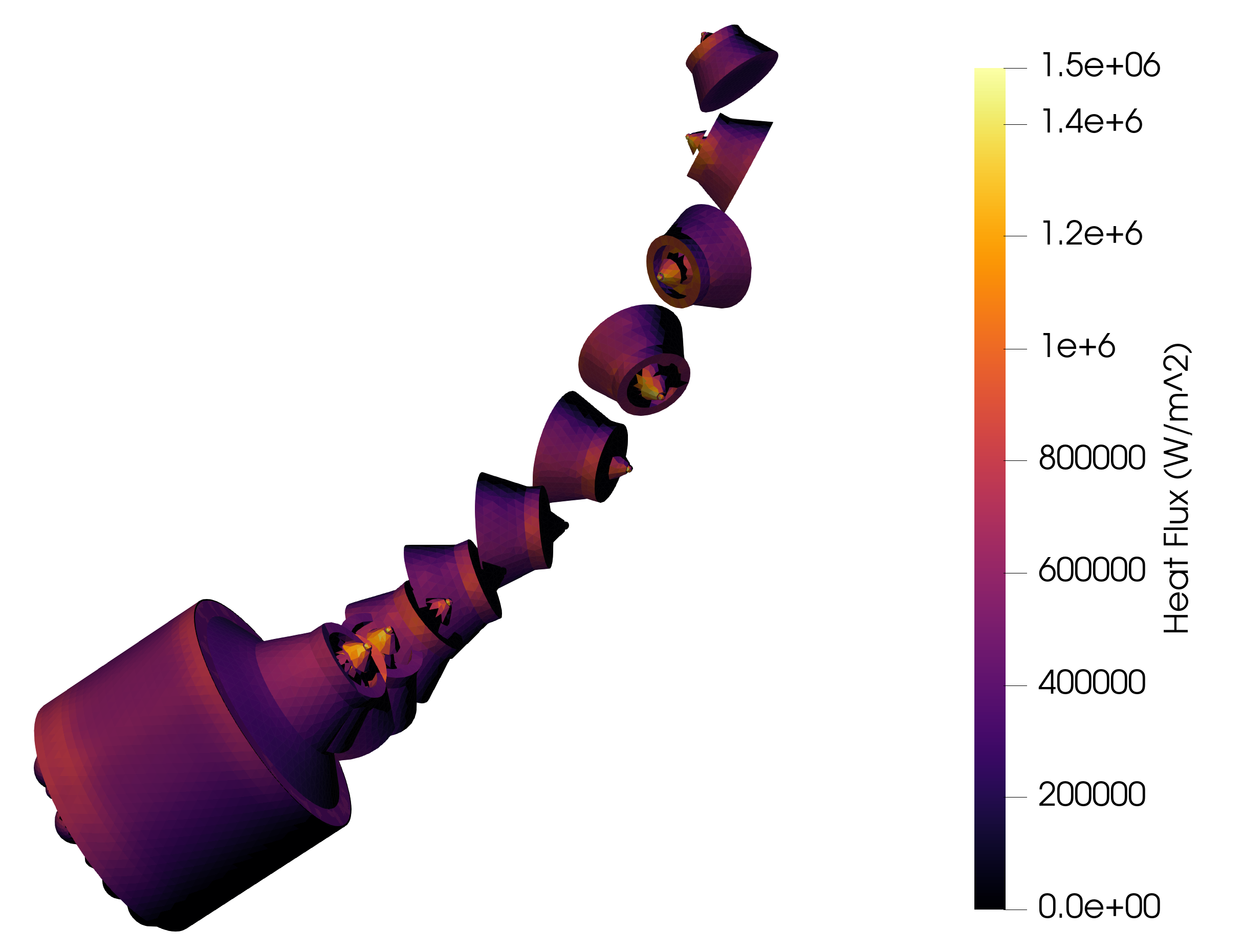}
    \caption{The ``cap'' of the equipped bay separating from the bay  ($\mathbf{\Delta t=0.32s}$, Bay frame)}
    \label{fig:captime}
\end{figure}
\subsection{REBR Behaviour}\label{ssec:Capsule}
In \autoref{fig:rebrtime} a forward-facing (towards the docking system) view of the equipped payload bay can be seen with stacked timesteps showing the chaotic bouncing behaviour of the REBR, contrasting the more conventional dispersion pattern displayed by the separation of the docking system and equipped bay ``cap'' \autoref{fig:captime}. The observed behaviour is emergent from TITAN's collision model as applied to internal motion in a cargo bay. This result is interesting when compared to the conjectured event log and data reported by Ailor\cite{ailor_reentry_2013}. The REBR's rate gyros become saturated \textit{before} any temperature increase is logged by the REBR's thermocouples. The event log suggests the REBR may be \textit{``bouncing''}, a behaviour that can be considered to be reproduced by this simulation. The hypothesis formed here echoes this proposal, specifically it is hypothesised that at $\sim 71.2$\unit{\kilo\metre} the REBR, and likely the housing that contained it, came free from its mounting in the ATV by mechanical failure but that it did not escape the cargo bay until significantly later in the re-entry. An additional flow phenomenon that was observed during simulation is that shadowed regions in the flow introduced by cavities such as the cargo bay are \textit{attractive}, in that other bodies require significant energy to escape them. According to the low-fidelity flow physical modelling performed here this is to be expected as the ray-tracing shadowing algorithm produces an adverse pressure gradient across the boundaries of such cavities that may or may not be representative of the true flow field. The fact that it is unknown to what extent this behaviour is simply an artifact of limited modelling should motivate further study of these phenomena using higher-fidelity methods.
\begin{figure*}[ht]
    \centering
    \begin{subfigure}{0.49\linewidth}
            \centering
            \includegraphics[width=\textwidth]{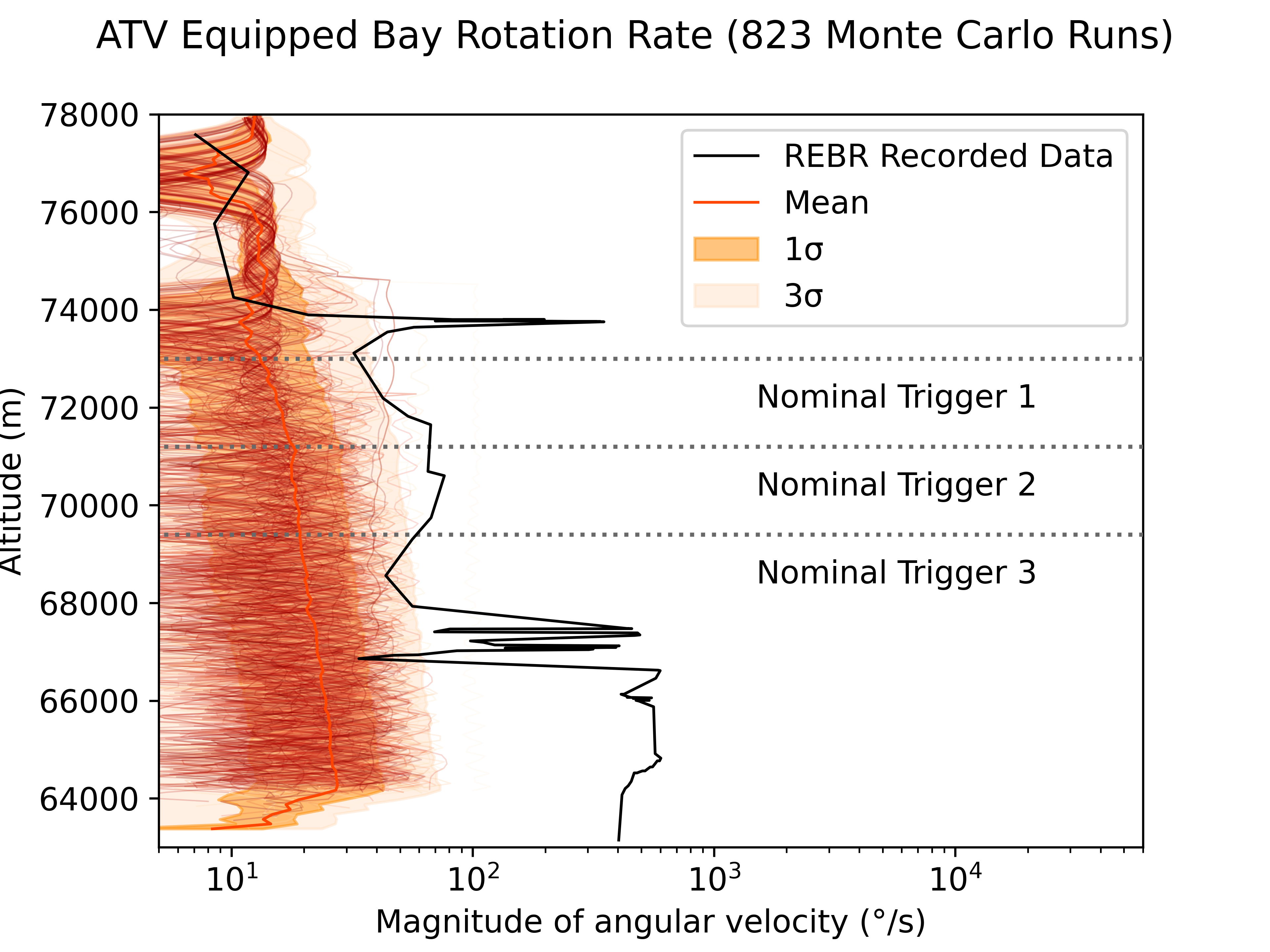}
            \caption{ATV equipped bay case}
            \label{subfig:bay_omega}
        \end{subfigure}
    \begin{subfigure}{0.49\linewidth}
            \centering
            \includegraphics[width=\textwidth]{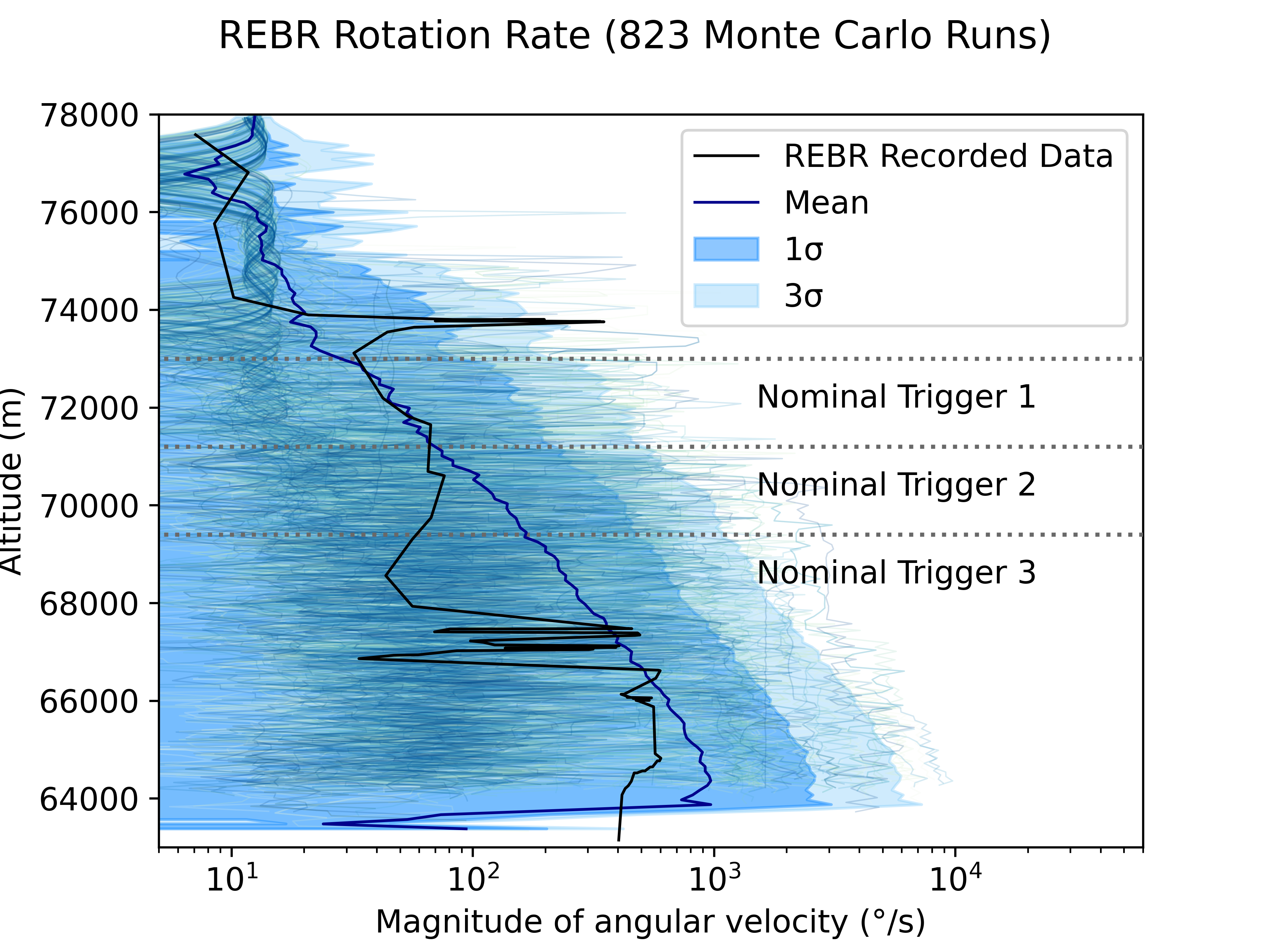}
            \caption{REBR case}
            \label{subfig:omega}
    \end{subfigure}
    \caption{Comparison of rotation rate magnitude of the ATV equipped bay case (\autoref{sub@subfig:bay_omega}) and the REBR case (\autoref{sub@subfig:omega})}
    \label{fig:omega}
\end{figure*}
\section{Stochastic Simulation of ATV Re-entry: Monte Carlo Campaign}\label{sec:uq}
Further to the initial deterministic exploration, the problem was approached from a stochastic perspective with a Monte Carlo Campaign. The simulation was perturbed from the nominal parameters through uncertainties applied to specific parameters of the problem recognised to have a substantial impact on the dynamics. 823 simulations were performed, each with a duration of $365$s of re-entry starting from the entry interface down to approximately 63km. 
\subsection{Uncertain Parameters}
\autoref{tab:ATV_uq} presents the uncertain parameters and their distributions. Normal distributions are applied to each event trigger with $\sigma=2$\unit{\kilo\metre} according to the European Space Agency's space debris mitigation guidelines for simulation\cite{noauthor_esa_nodate}. An initial attitude is selected via Haar-Uniform Quaternion sampling that results in a uniform distribution in rotation space. Angular velocity is also randomly perturbed as the ATV was entered into an unknown tumbling state for re-entry. Finally an atmospheric density multiplicative factor is applied.
\begin{table*}[h]
    \caption{ATV3 Uncertain Parameters}
    \label{tab:ATV_uq}
    \centering
    \begin{tabular*}{\linewidth}{llll}
        \toprule
        \textbf{Parameter} & \textbf{Distribution} & \textbf{Representation} & \\
        \midrule
        Propulsion Bay separation & Normal & $\mu=73$\unit{\kilo\metre},& $\sigma=2$\unit{\kilo\metre}\\
        REBR Separation & Normal & $\mu=71.2$\unit{\kilo\metre},& $\sigma=2$\unit{\kilo\metre}\\
        Equipped Bay Cap separation & Normal & $\mu=69.4$\unit{\kilo\metre},& $\sigma=2$\unit{\kilo\metre}\\
        Initial Attitude & Haar-Uniform Quaternion & N/A&N/A\\
        Initial Rotation Rate & Trivariate Normal & $\mu=(0, 10.0, 0)$\unit{\degree\per\second},& $\textbf{C}=3^2\cdot\mathbf{I}$\unit{\degree\per\second}\\
        Atmospheric Density Factor & Uniform & $l=0.8$,& $u=1.2$\\
        \bottomrule
    \end{tabular*}
\end{table*}
\subsection{Results}
The very high mass differential between the REBR and the ATV mean that in the freeflying case any influence of the REBR upon the motion of the ATV can be safely neglected as negligible. Thus here the motion of the ATV after the separation of the REBR is considered to be equivalent, for all intents and purposes, to the motion in the attached case. Recalling that rigid bodies must have everywhere constant angular velocity the collected results can be decomposed into ``whole'' and ``bouncing'' cases. Specifically in the results presented below those labelled as \textit{ATV Equipped Bay} represent the motion of the container vehicle bay and REBR as a single rigid body whereas those labelled \textit{REBR} represent the motion of the REBR as a separated body. Therefore the bouncing hypothesis can be interrogated through comparison of the magnitudes of angular velocity recorded by the ATV equipped payload bay and the separated REBR. \autoref{fig:omega} shows the rotation rates of these components plotted against altitude with the recorded angular velocity of the REBR shown in black. It can be seen in \autoref{subfig:bay_omega} that the recorded value for angular velocity is approximately an order of magnitude above the $3\sigma$ boundary of the equipped bay's rotation, thus it can be reasoned that the situation where the REBR remains attached at altitudes below 72\unit{\kilo\metre} is unlikely. In contrast it can be seen in \autoref{subfig:omega} that the recorded data is encompassed by the $1\sigma$ region of the bouncing case, lending significant evidence in support of the hypothesis of the detachment of the recorder from the cargo bay being the reason for the measured rotation rates. It is worth noting that the data recovered from the REBR exceeded its gyro rate limits of 300 \unit{\degree\per\second}, so the actual behaviour of the recorder could have indeed featured larger rotational rates than were measured.

\begin{figure*}[p]
    \centering
    \begin{subfigure}{0.9\linewidth}
            \centering
            \includegraphics[width=\textwidth]{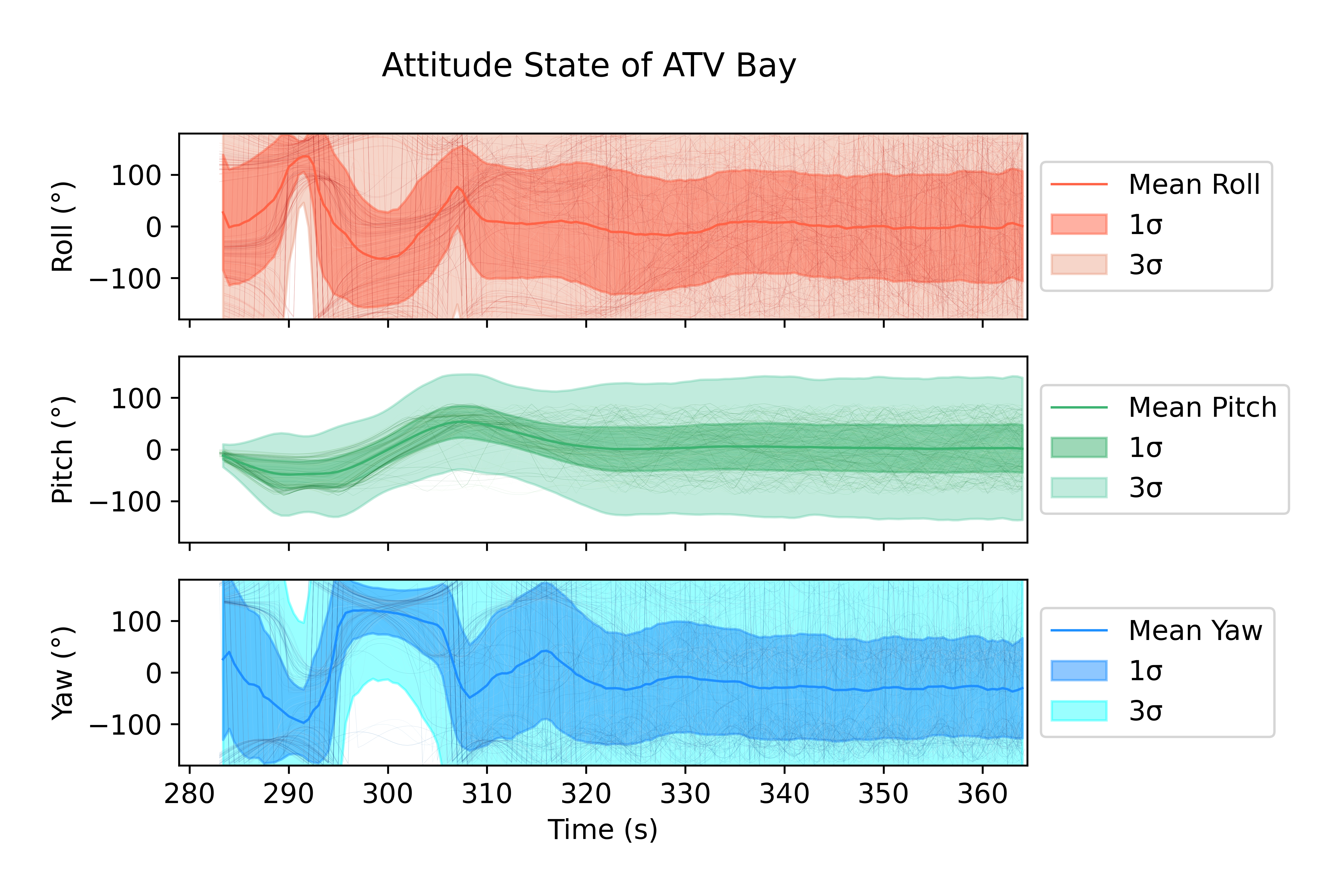}
            \caption{ATV equipped bay case}
            \label{subfig:bay_rpy}
        \end{subfigure}\\
    \begin{subfigure}{0.9\linewidth}
            \centering
            \includegraphics[width=\textwidth]{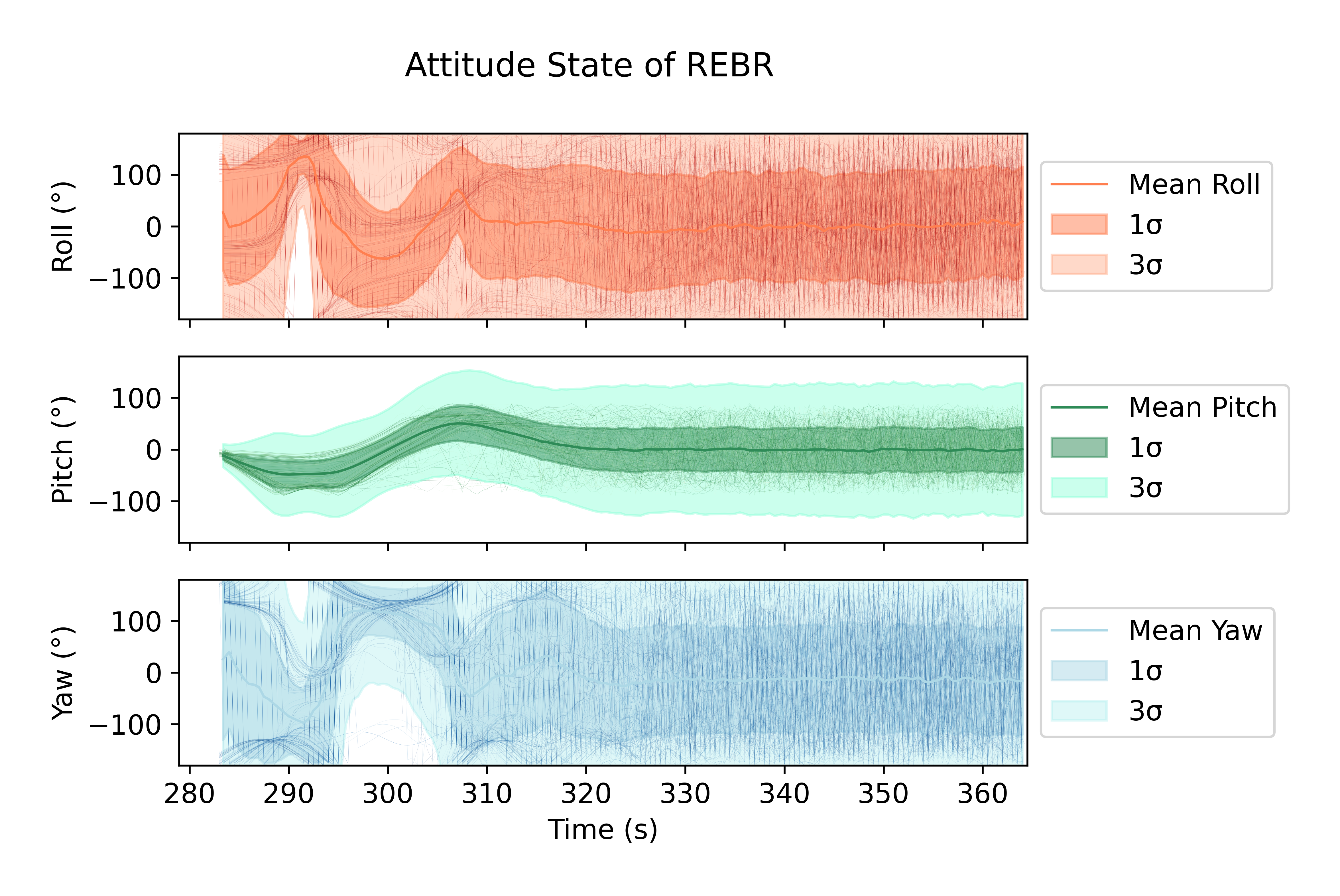}
            \caption{REBR case}
            \label{subfig:rebr_rpy}
    \end{subfigure}
    \caption{Comparison of attitude state of the ATV equipped bay case(\autoref{sub@subfig:bay_rpy}) and the REBR case (\autoref{sub@subfig:rebr_rpy})}
    \label{fig:rpy}
\end{figure*}

\begin{figure*}[p]
    \centering
    \begin{subfigure}{0.9\linewidth}
            \centering
            \includegraphics[width=\textwidth]{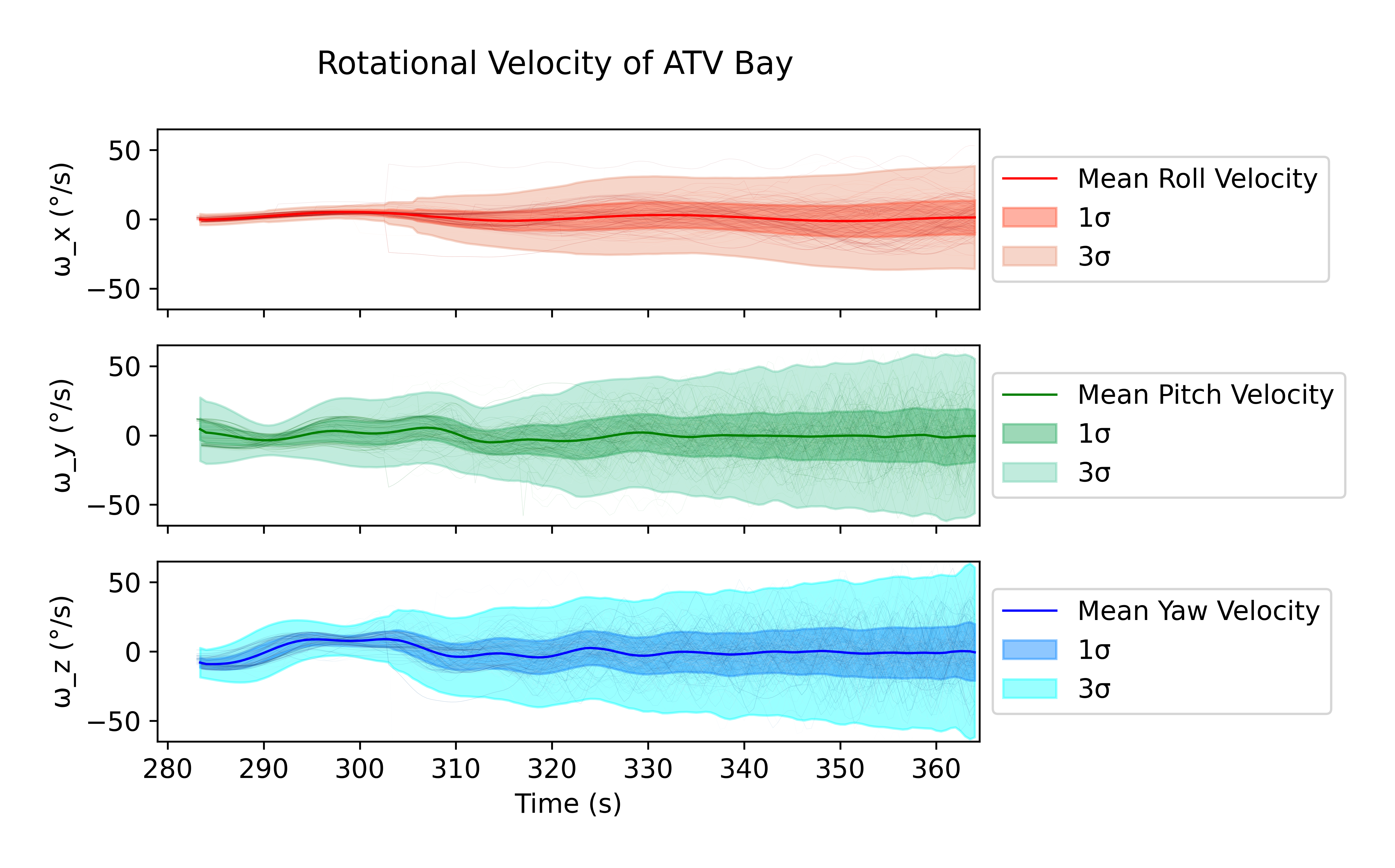}
            \caption{ATV equipped bay case}
            \label{subfig:bay_vrpy}
        \end{subfigure}\\
    \begin{subfigure}{0.9\linewidth}
            \centering
            \includegraphics[width=\textwidth]{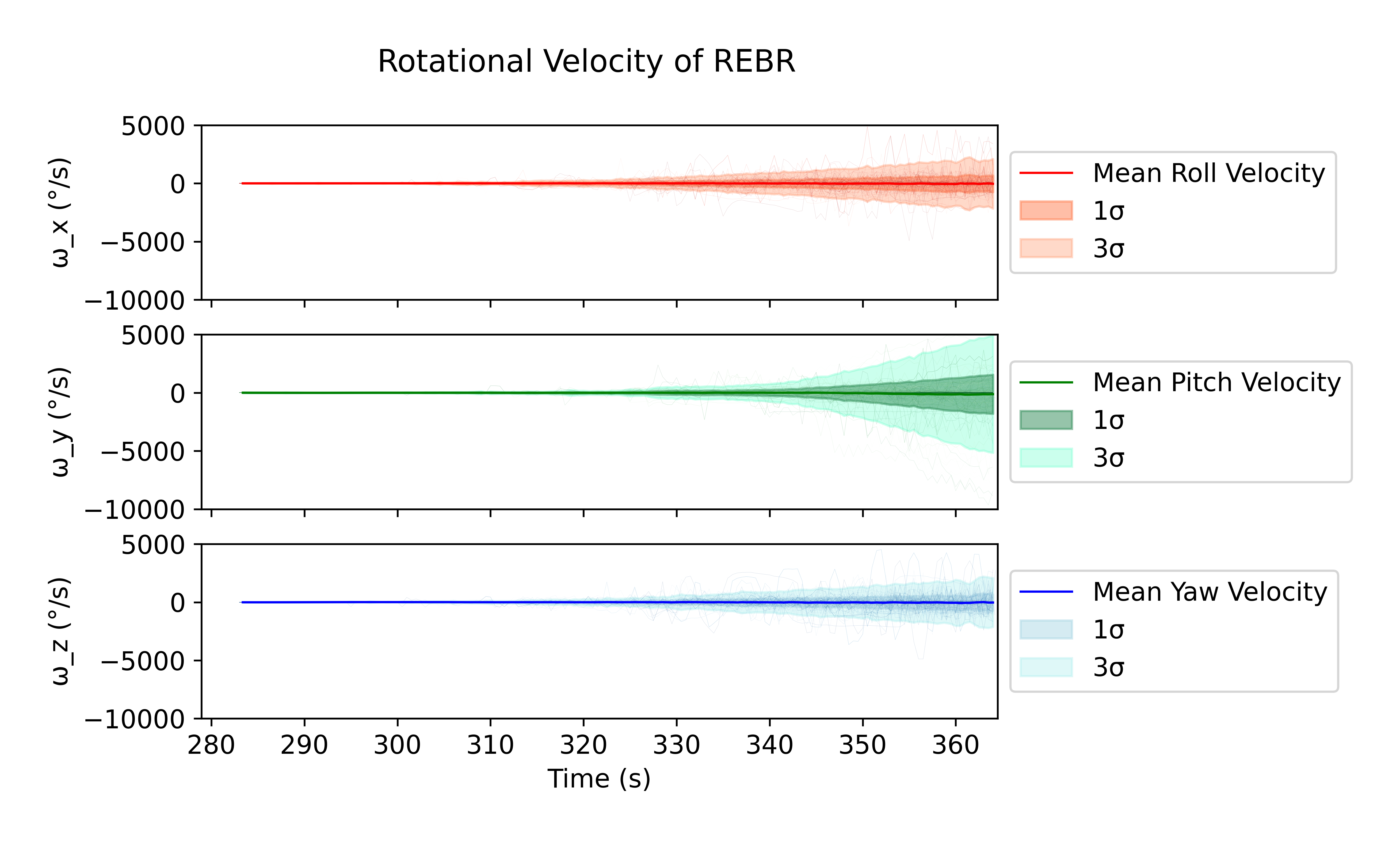}
            \caption{REBR case}
            \label{subfig:rebr_vrpy}
    \end{subfigure}
    \caption{Comparison of per-axis rotation rates of the ATV equipped bay case(\autoref{sub@subfig:bay_vrpy}) and the REBR case (\autoref{sub@subfig:rebr_vrpy})}
    \label{fig:vrpy}
\end{figure*}

\begin{figure*}[p]

    \begin{subfigure}{0.49\linewidth}
            \centering
            \includegraphics[width=\textwidth]{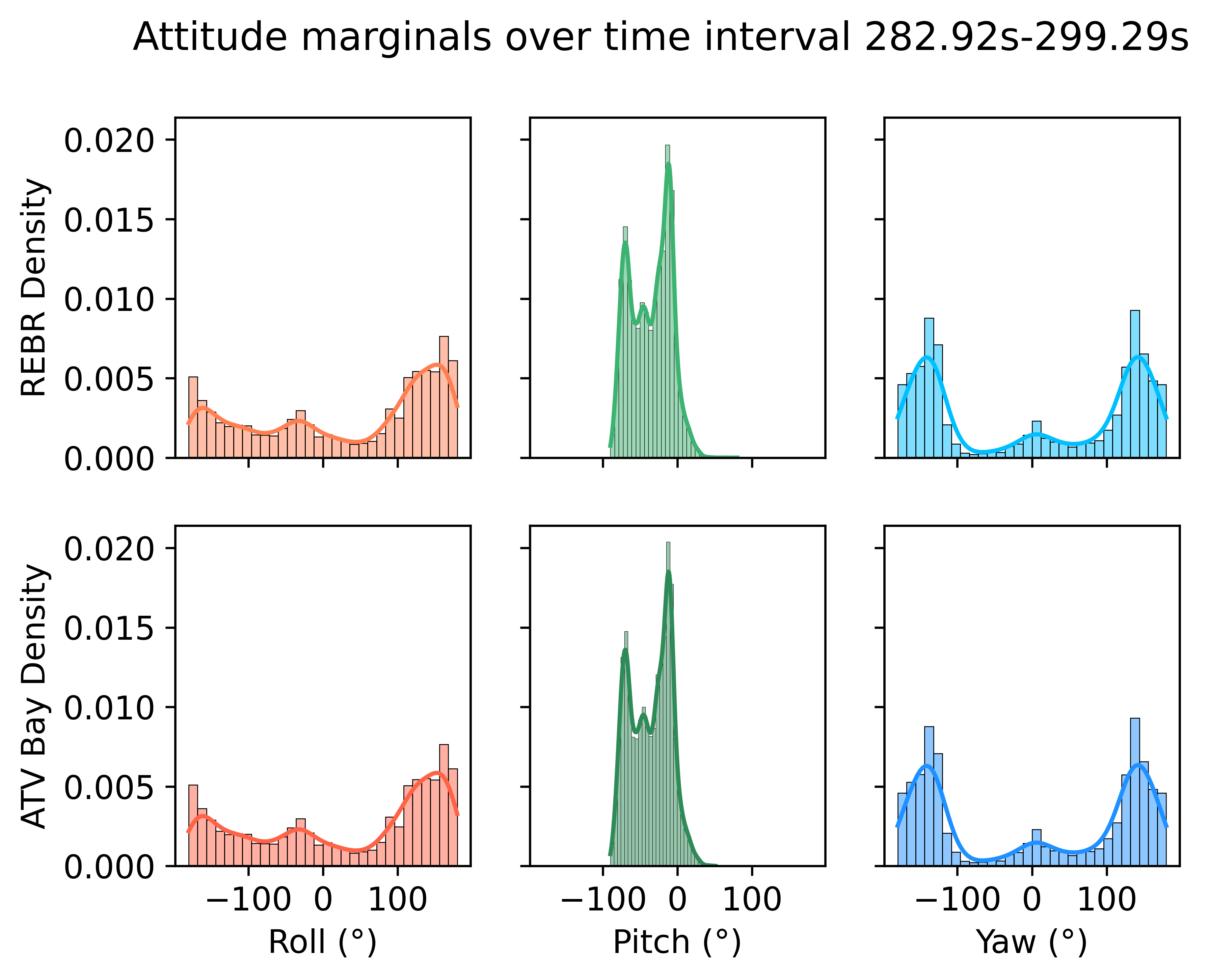}
            \caption{$282.92<t<299.29$}
            \label{fig:r282.92}
        \end{subfigure}
    \begin{subfigure}{0.49\linewidth}
            \centering
            \includegraphics[width=\textwidth]{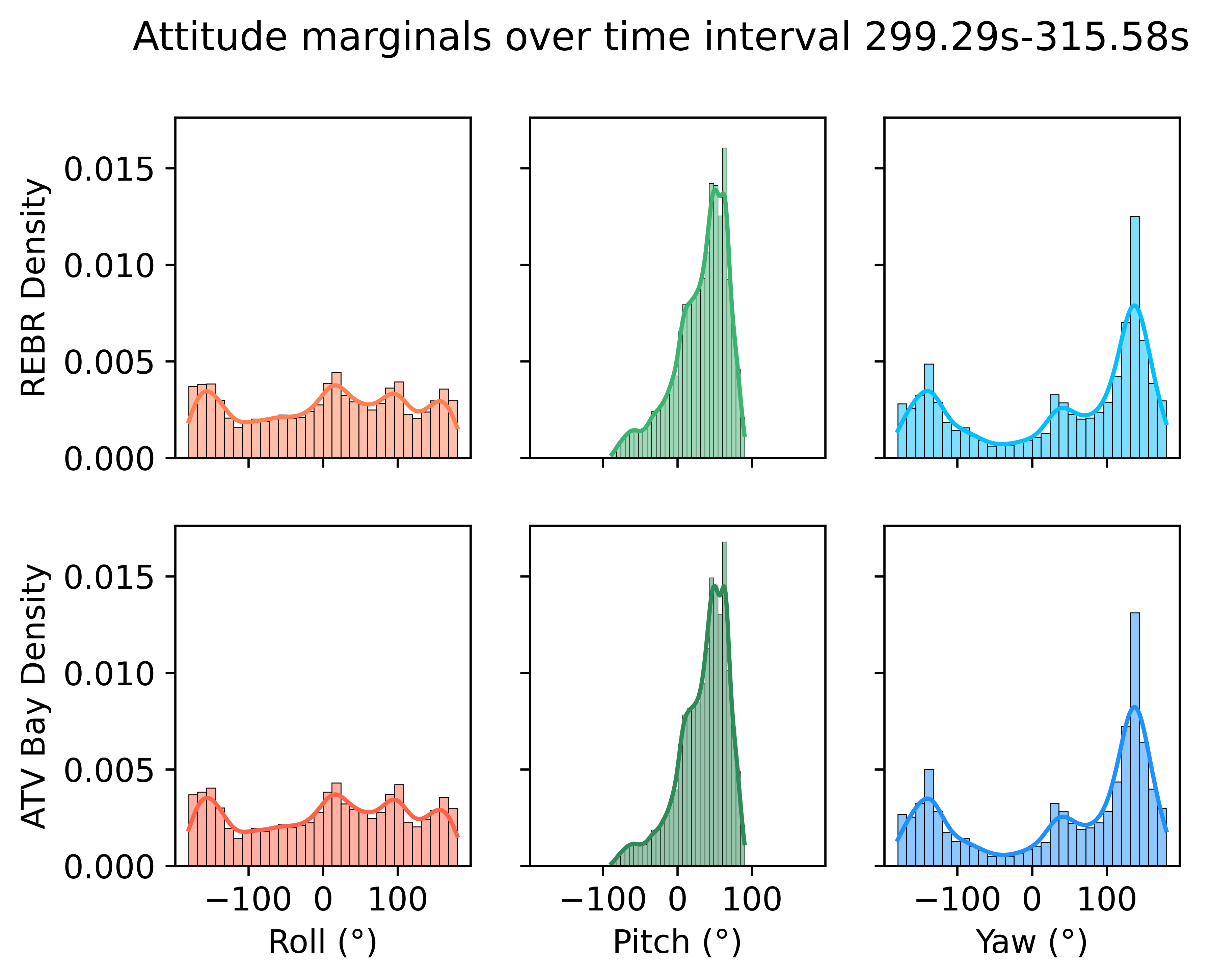}
            \caption{$299.29<t<315.58$}
            \label{fig:r299.29}
    \end{subfigure}
    \begin{subfigure}{0.49\linewidth}
            \centering
            \includegraphics[width=\textwidth]{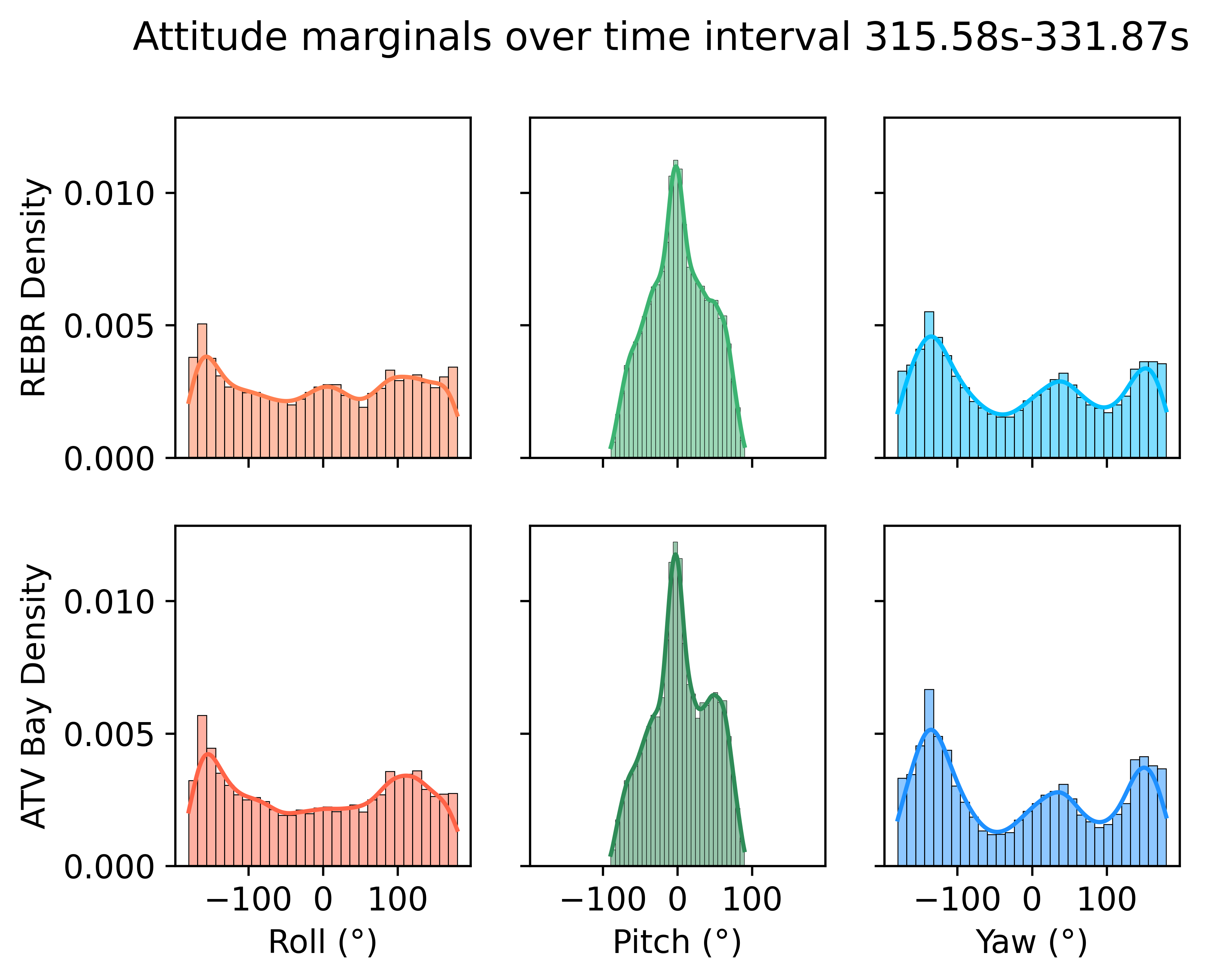}
            \caption{$315.58<t<331.87$}
            \label{fig:r315.58}
    \end{subfigure}
    \begin{subfigure}{0.49\linewidth}
            \centering
            \includegraphics[width=\textwidth]{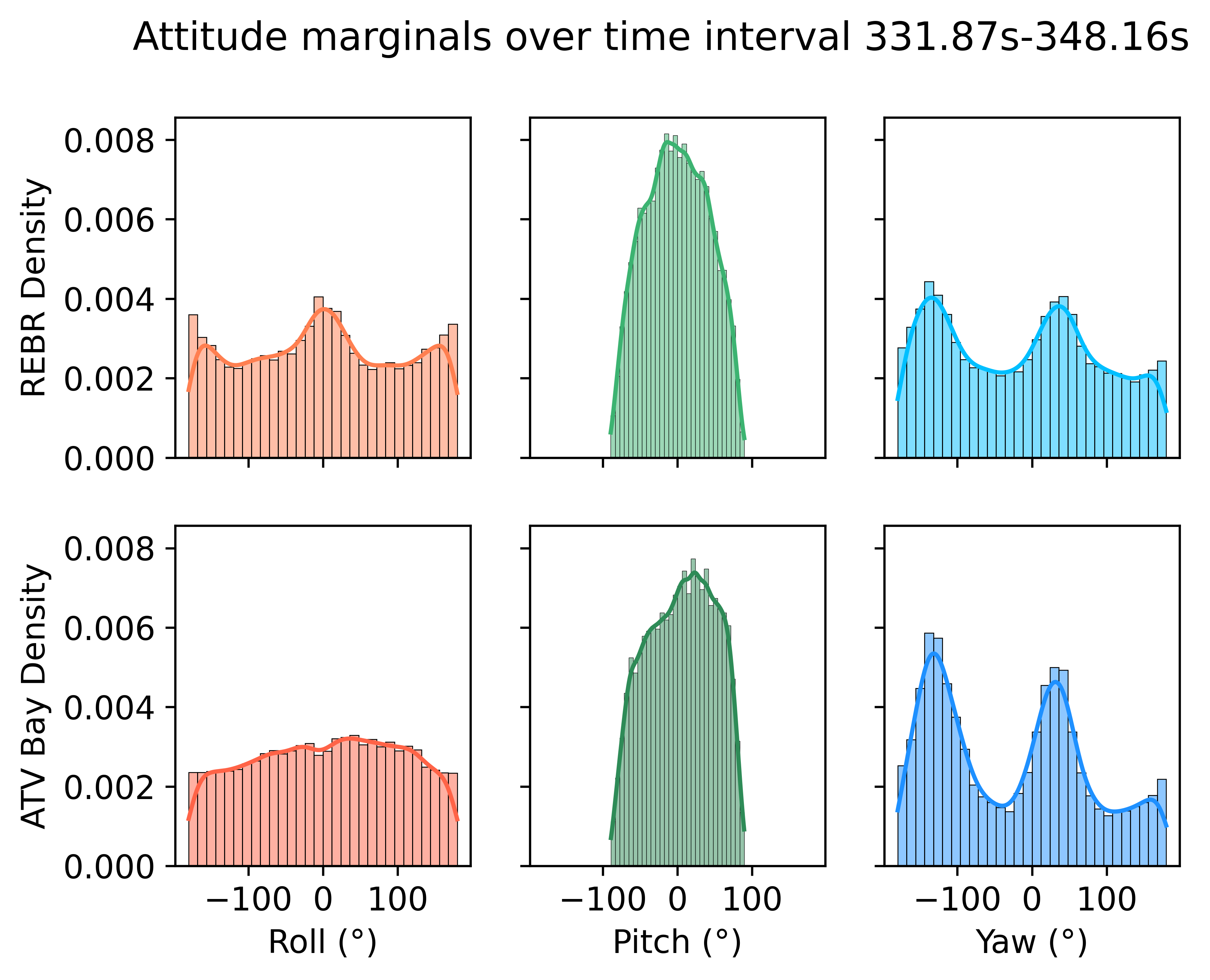}
            \caption{$331.87<t<348.16$}
            \label{fig:r331.87}
    \end{subfigure}
    \begin{subfigure}{0.49\linewidth}
            \centering
            \includegraphics[width=\textwidth]{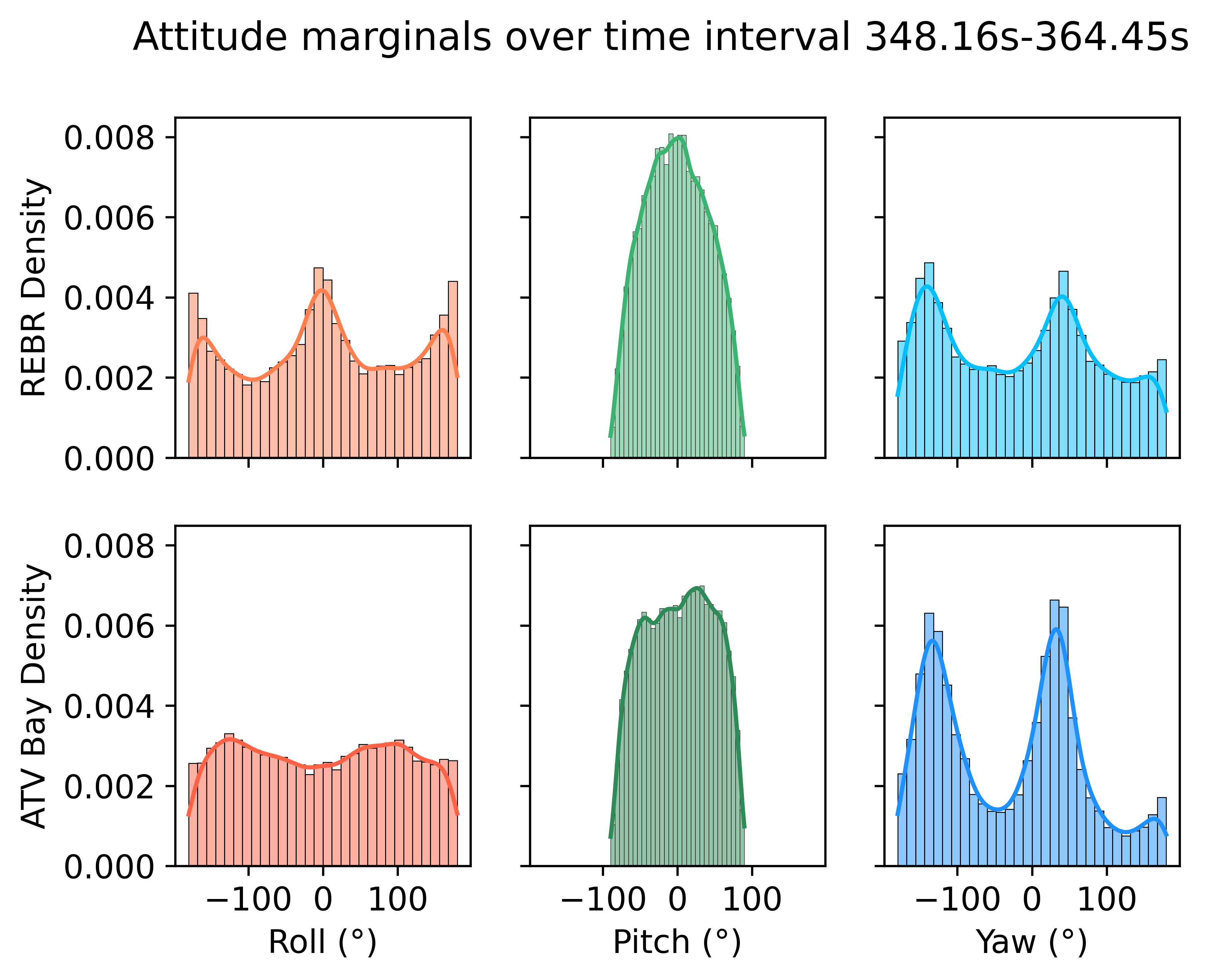}
            \caption{$348.16<t<364.45$}
            \label{fig:r348.16}
    \end{subfigure}
    \caption{Time-binned attitude state marginals of the ATV equipped bay case and REBR case}
    \label{fig:rolcompfig}
\end{figure*}

\begin{figure*}[h]
    \begin{subfigure}{0.49\linewidth}
            \centering
            \includegraphics[width=\textwidth]{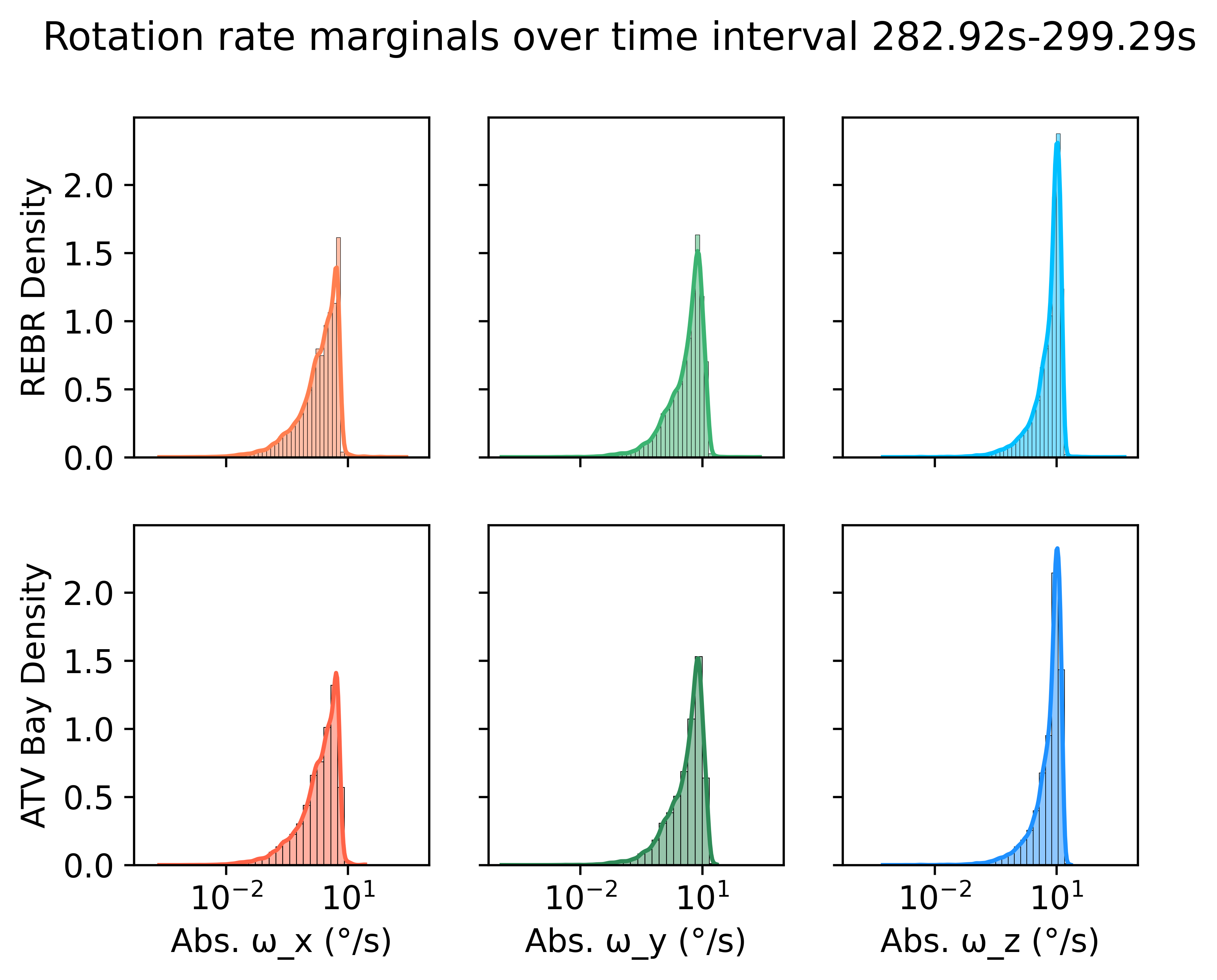}
            \caption{$282.92<t<299.29$}
            \label{fig:vr282.92}
        \end{subfigure}
    \begin{subfigure}{0.49\linewidth}
            \centering
            \includegraphics[width=\textwidth]{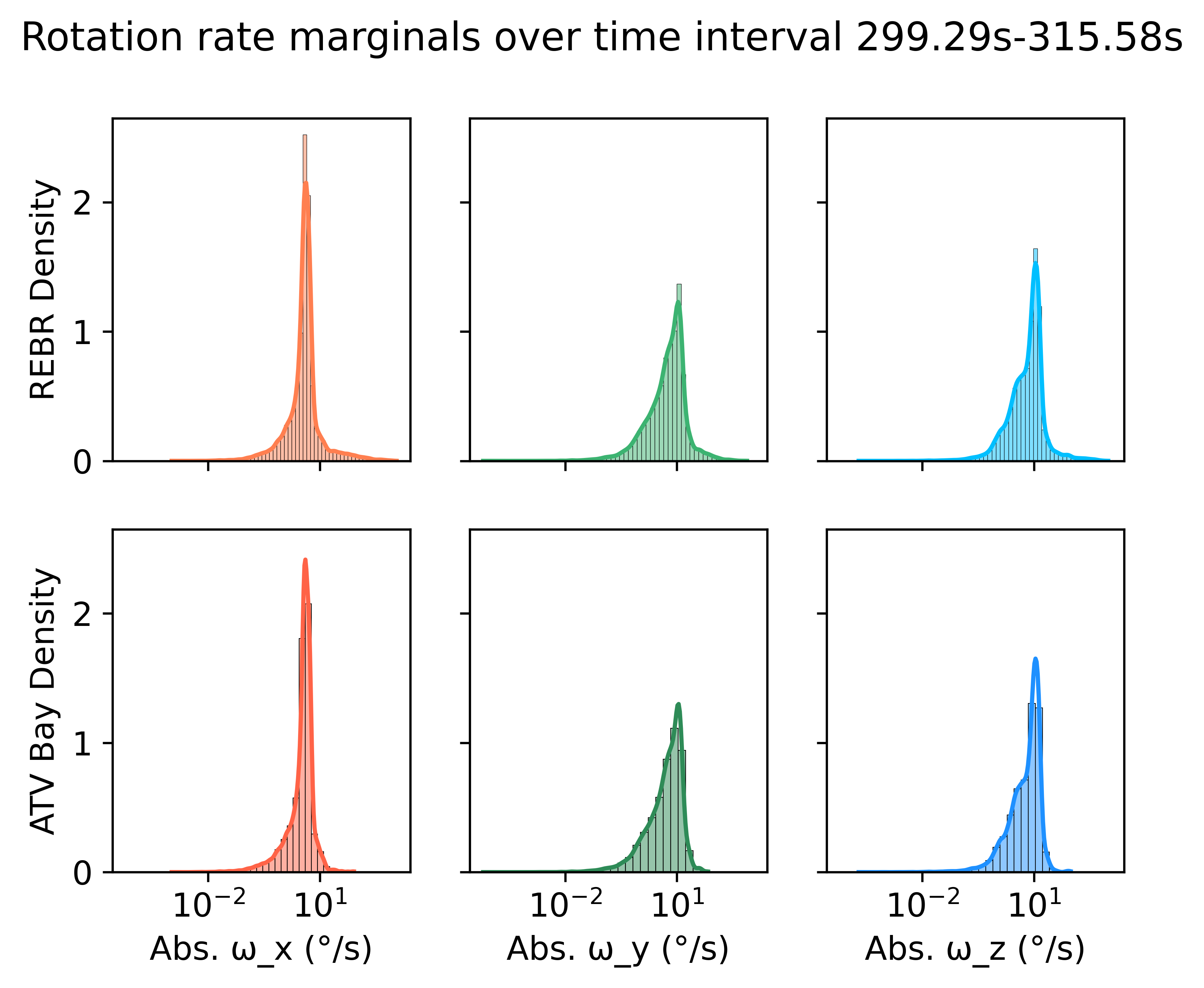}
            \caption{$299.29<t<315.58$}
            \label{fig:vr299.29}
    \end{subfigure}
    \begin{subfigure}{0.49\linewidth}
            \centering
            \includegraphics[width=\textwidth]{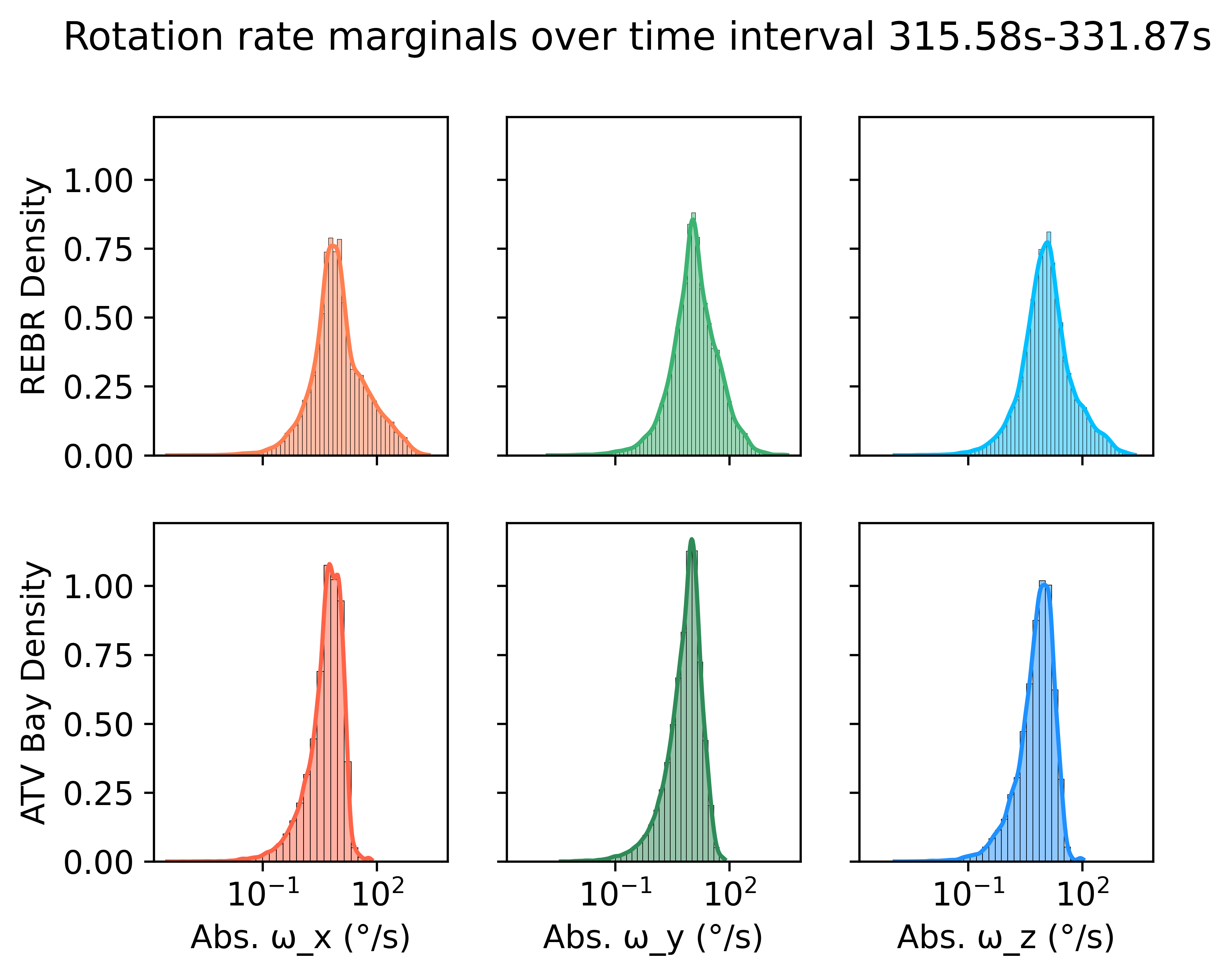}
            \caption{$315.58<t<331.87$}
            \label{fig:vr315.58}
    \end{subfigure}
    \begin{subfigure}{0.49\linewidth}
            \centering
            \includegraphics[width=\textwidth]{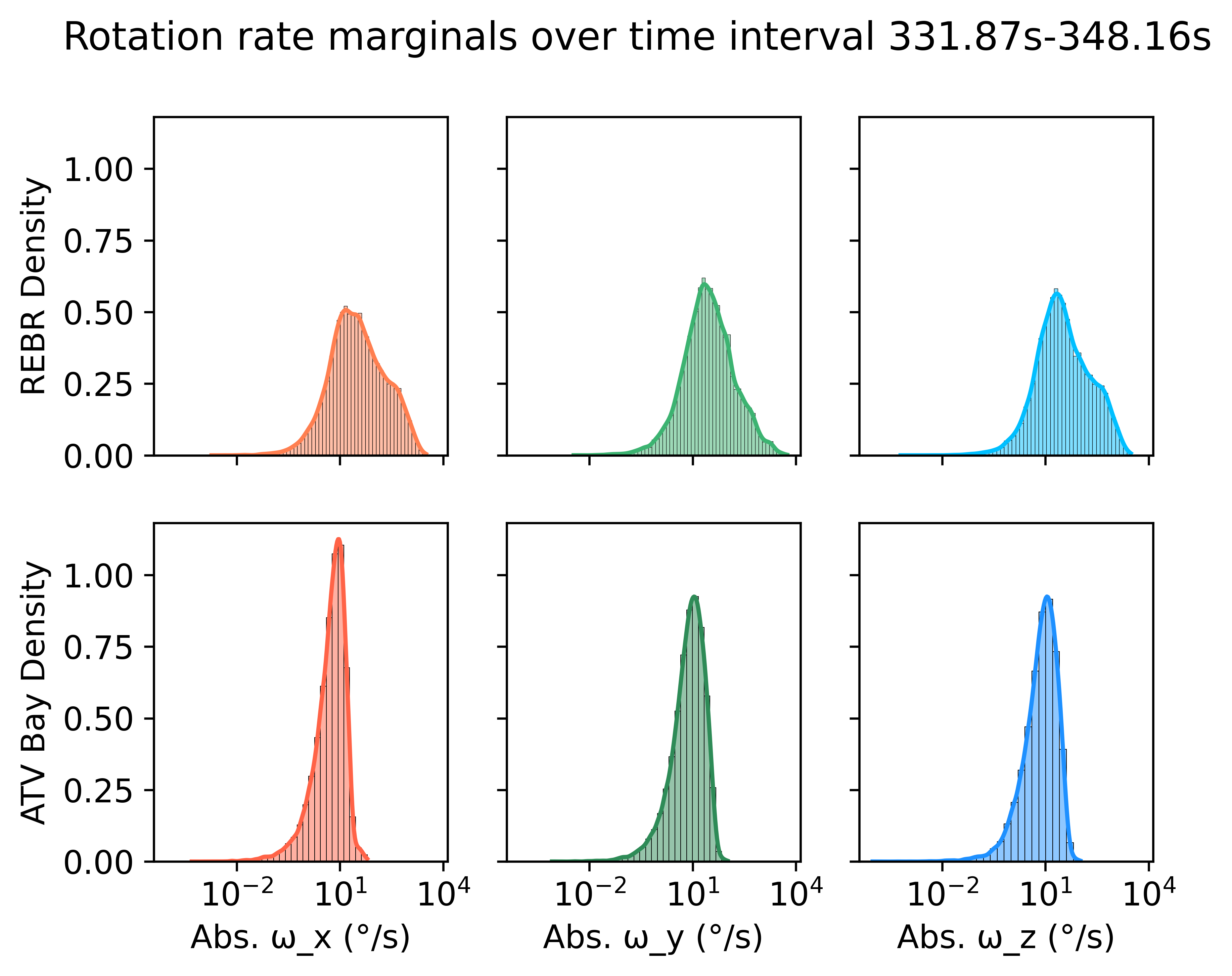}
            \caption{$331.87<t<348.16$}
            \label{fig:vr331.87}
    \end{subfigure}
    \begin{subfigure}{0.49\linewidth}
            \centering
            \includegraphics[width=\textwidth]{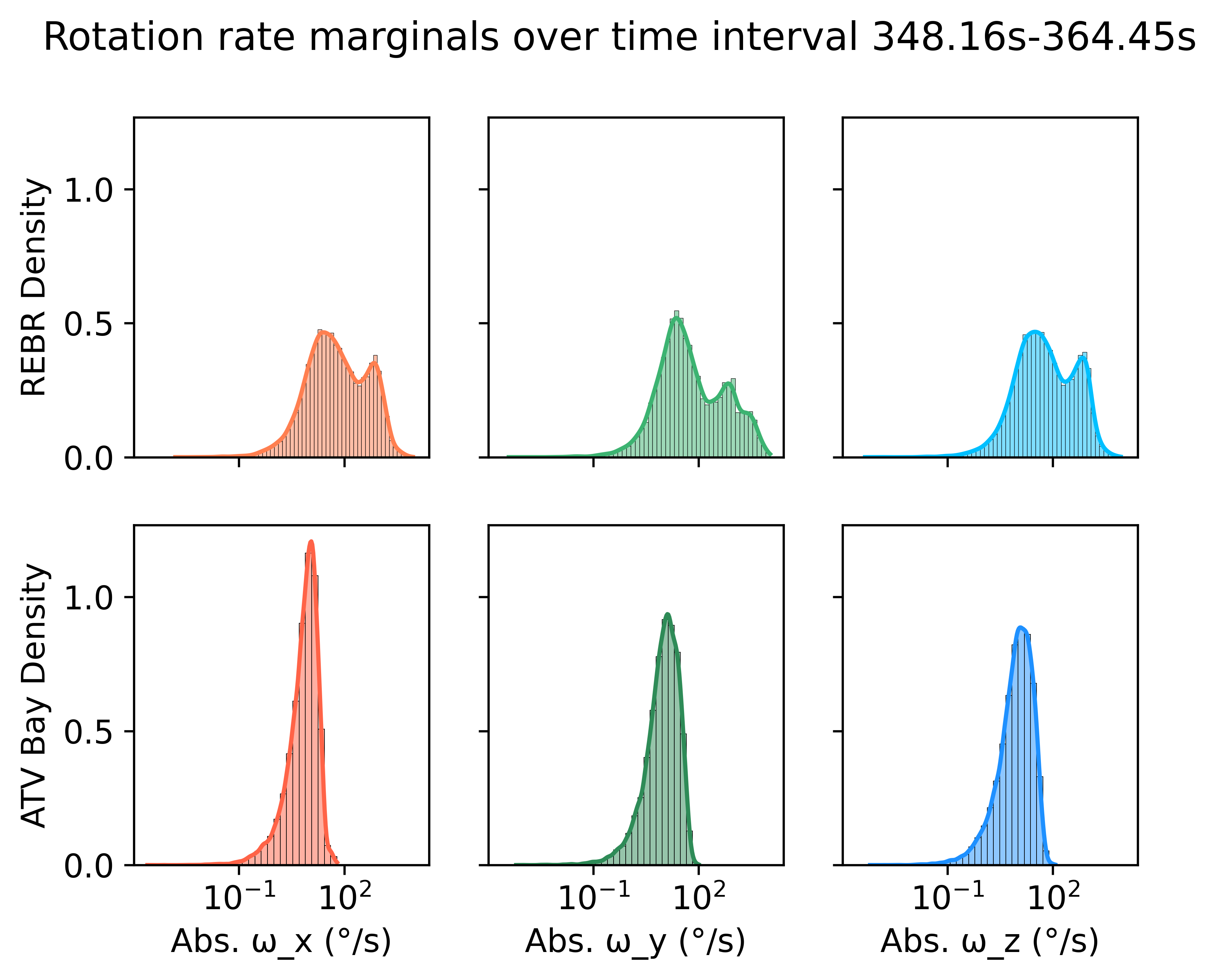}
            \caption{$348.16<t<364.45$}
            \label{fig:vr348.16}
    \end{subfigure}
    \caption{Time-binned angular velocity marginals of the ATV equipped bay case and the REBR case}
    \label{fig:vrolcompfig}
\end{figure*}

By inspecting the per-axis attitude (\autoref{fig:rpy}) it is evident that the state of the bay and REBR do not necessarily differ greatly in terms of attitude at any given moment of the re-entry. In addition, an evolution of attitude towards a broadly ``space-filling'' random tumbling state can be observed. 
However \autoref{fig:vrpy} shows significant departure in the magnitudes of rotational velocity, as previously mentioned. 
By time-binning attitude (\autoref{fig:rolcompfig}) and angular velocity (\autoref{fig:vrolcompfig}) it can be seen that the attitude state and velocity of the REBR gradually diverges from the bay into a bimodal distribution which can be inferred to be the split between cases where the REBR has escaped the bay and where it is still bouncing internally.

\begin{figure*}[htp]
    \begin{subfigure}{0.47\linewidth}
        \centering
        \includegraphics[width=0.95\textwidth]{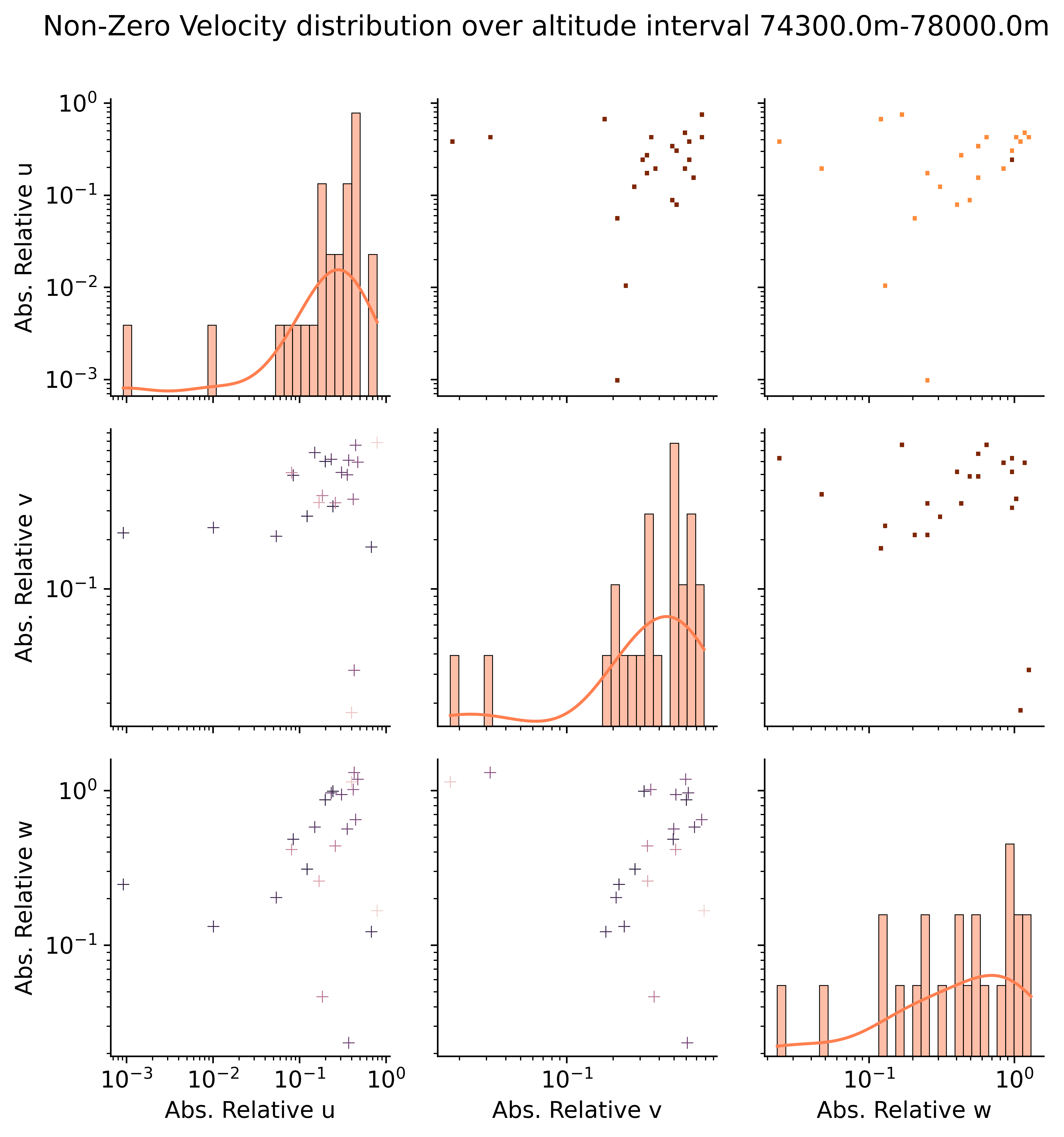}
        \caption{}
        \label{fig:velpair1}
    \end{subfigure}\hfill
    \begin{subfigure}{0.47\linewidth}
        \centering
        \includegraphics[width=0.95\textwidth]{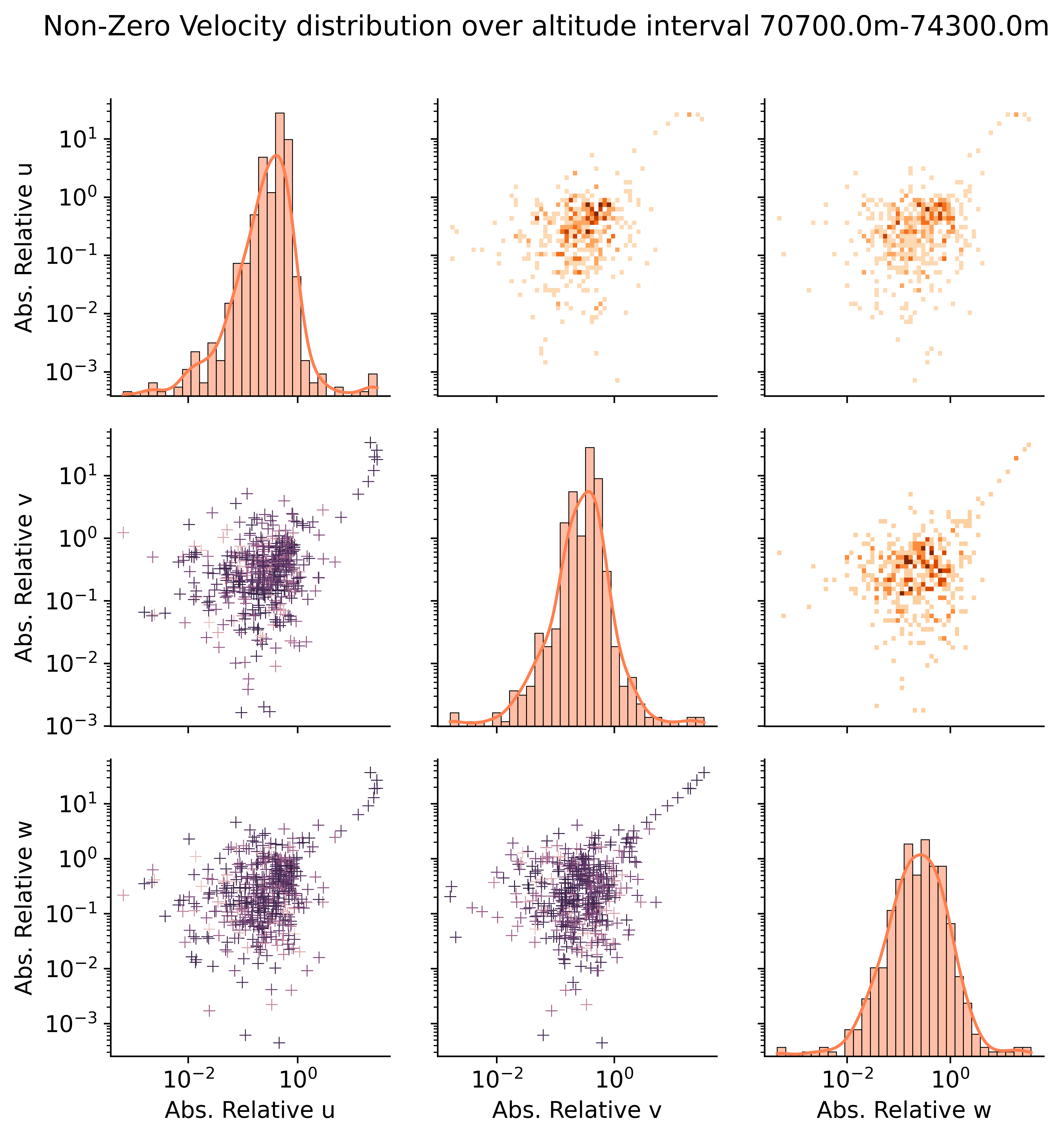}
        \caption{}
        \label{fig:velpair2}
    \end{subfigure}
    \begin{subfigure}{0.47\linewidth}
        \centering
        \includegraphics[width=0.95\textwidth]{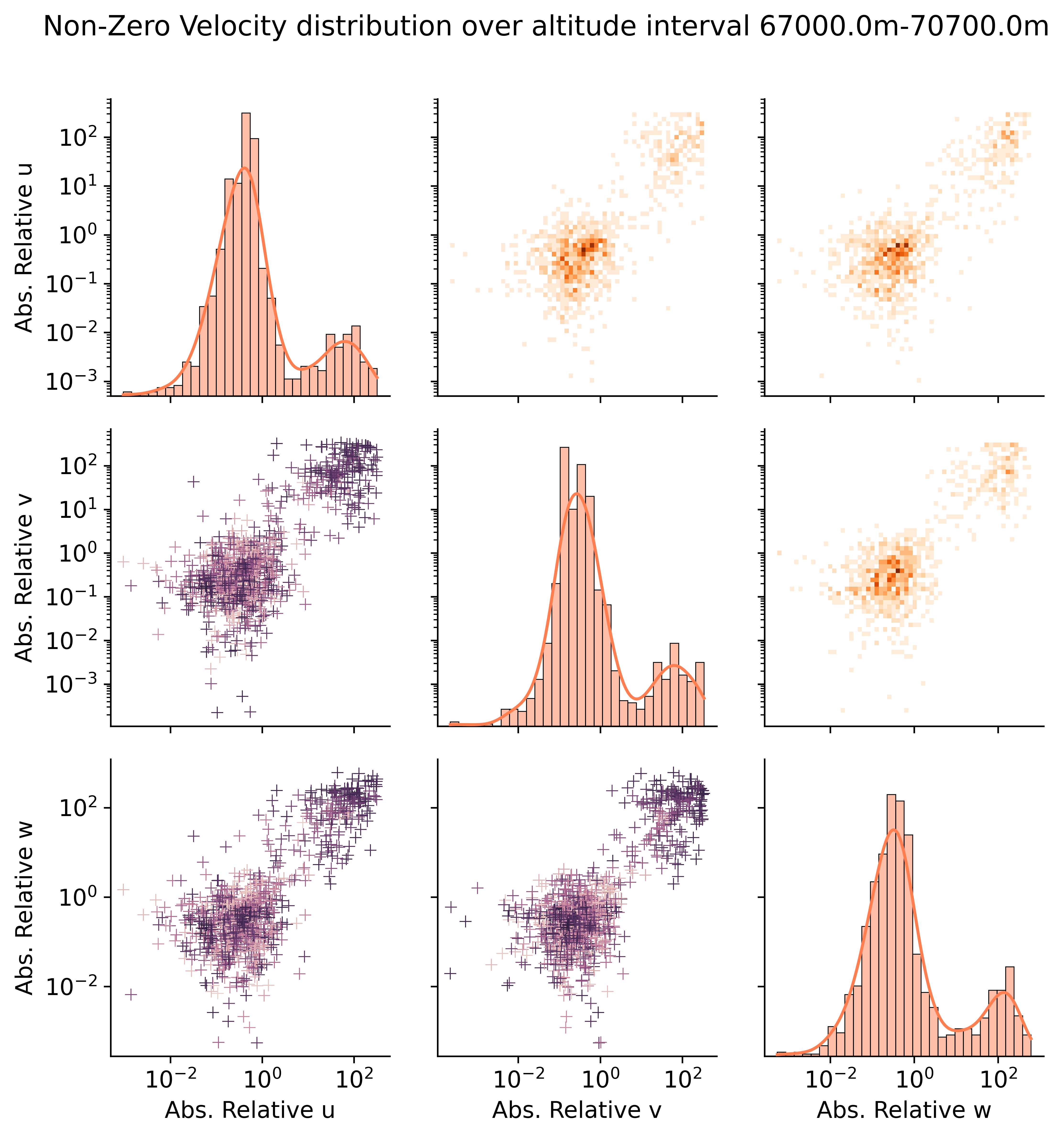}
        \caption{}
        \label{fig:velpair3}
    \end{subfigure}\hfill
    \begin{subfigure}{0.47\linewidth}
        \centering
        \includegraphics[width=0.95\textwidth]{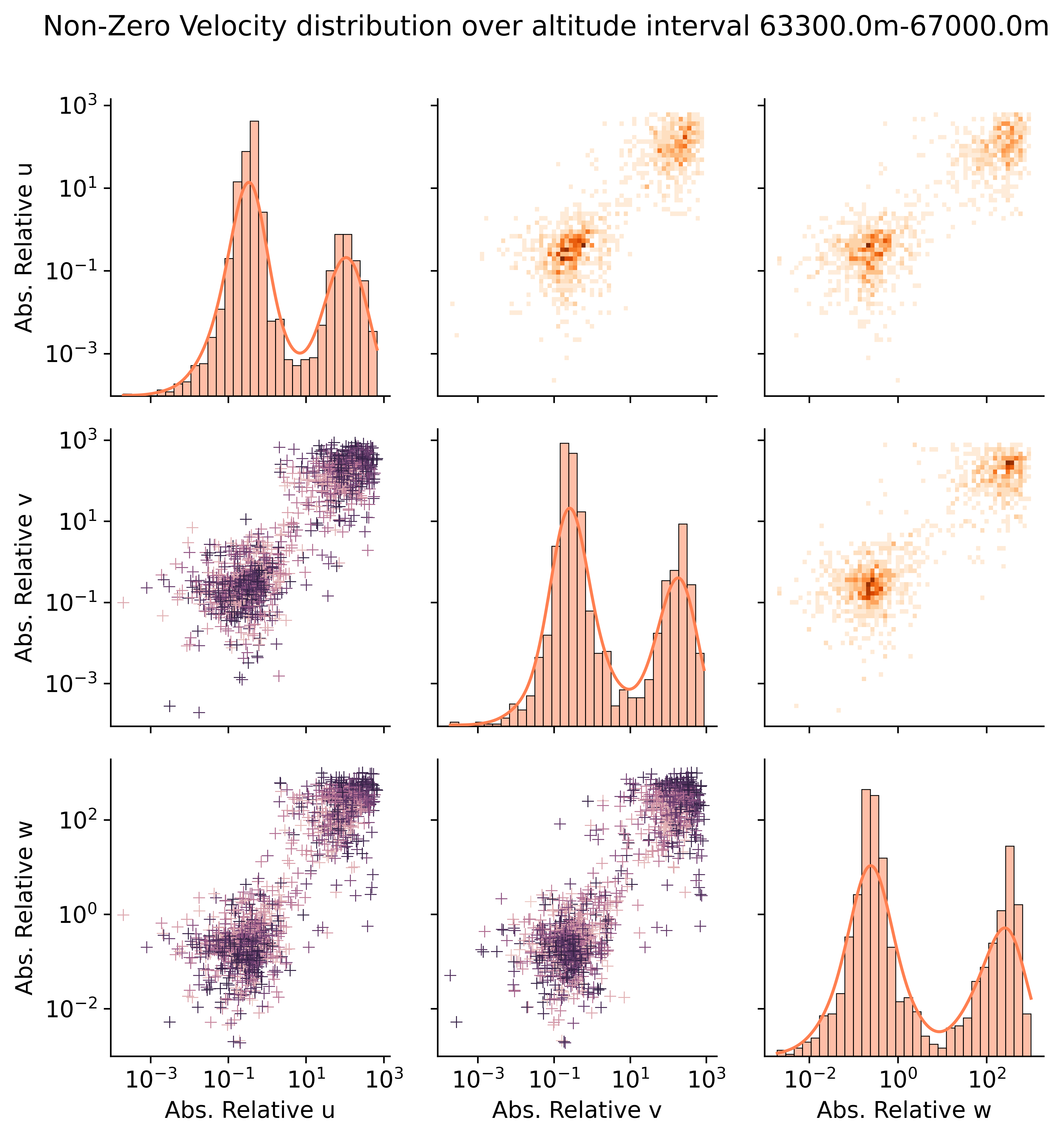}
        \caption{}
        \label{fig:velpair4}
    \end{subfigure}
    \caption{Log Relative velocity of the REBR and ATV Equipped Bay binned by altitude, colour indicates time}
    \label{fig:vel_pair}
\end{figure*}
\begin{figure}[h]
    \centering
    \includegraphics[width=0.75\textwidth]{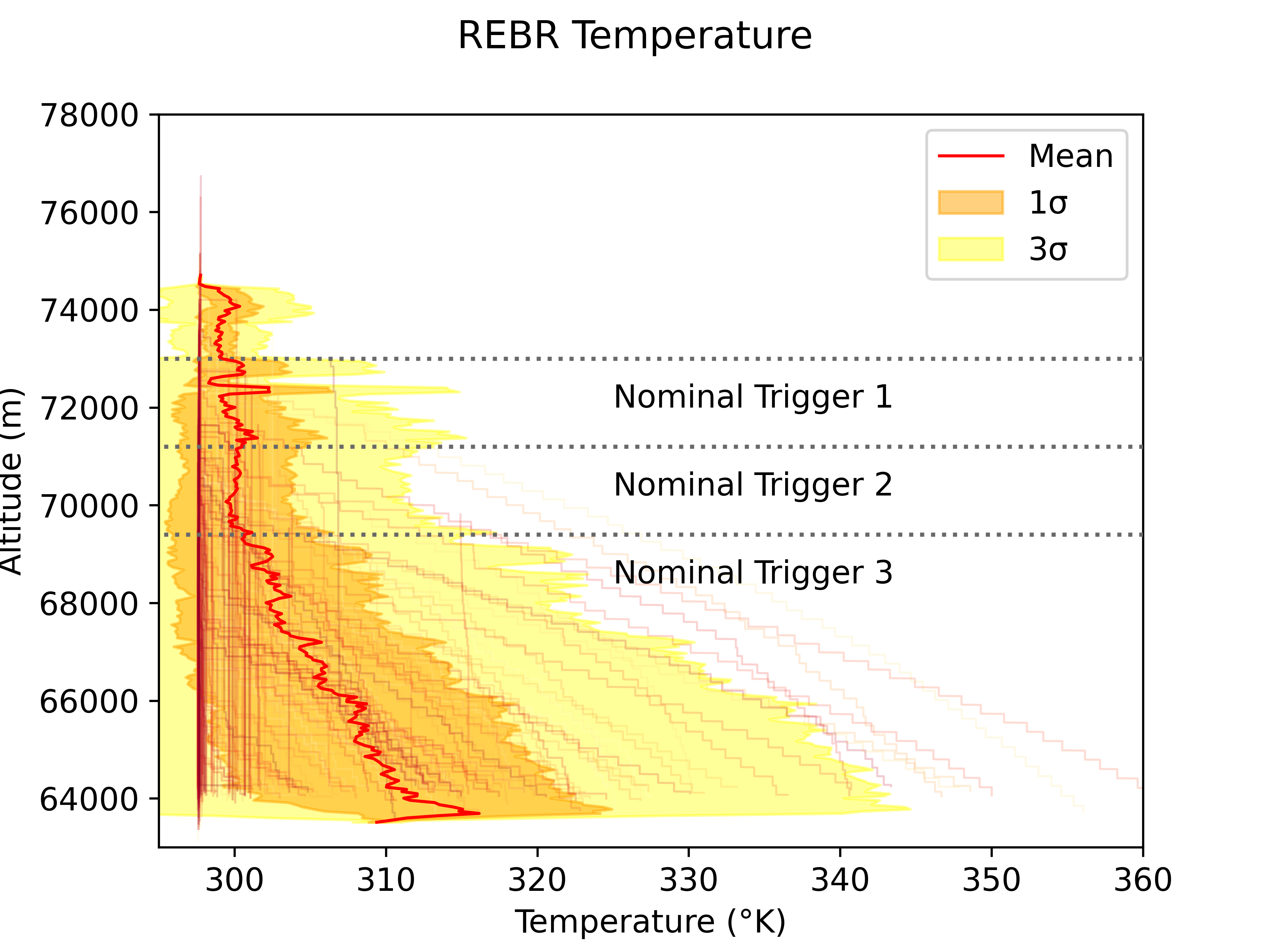}
    \caption{Temperature of the REBR}
    \label{fig:temp}
\end{figure}
\begin{figure}[h]
        \centering
        \includegraphics[width=0.75\textwidth]{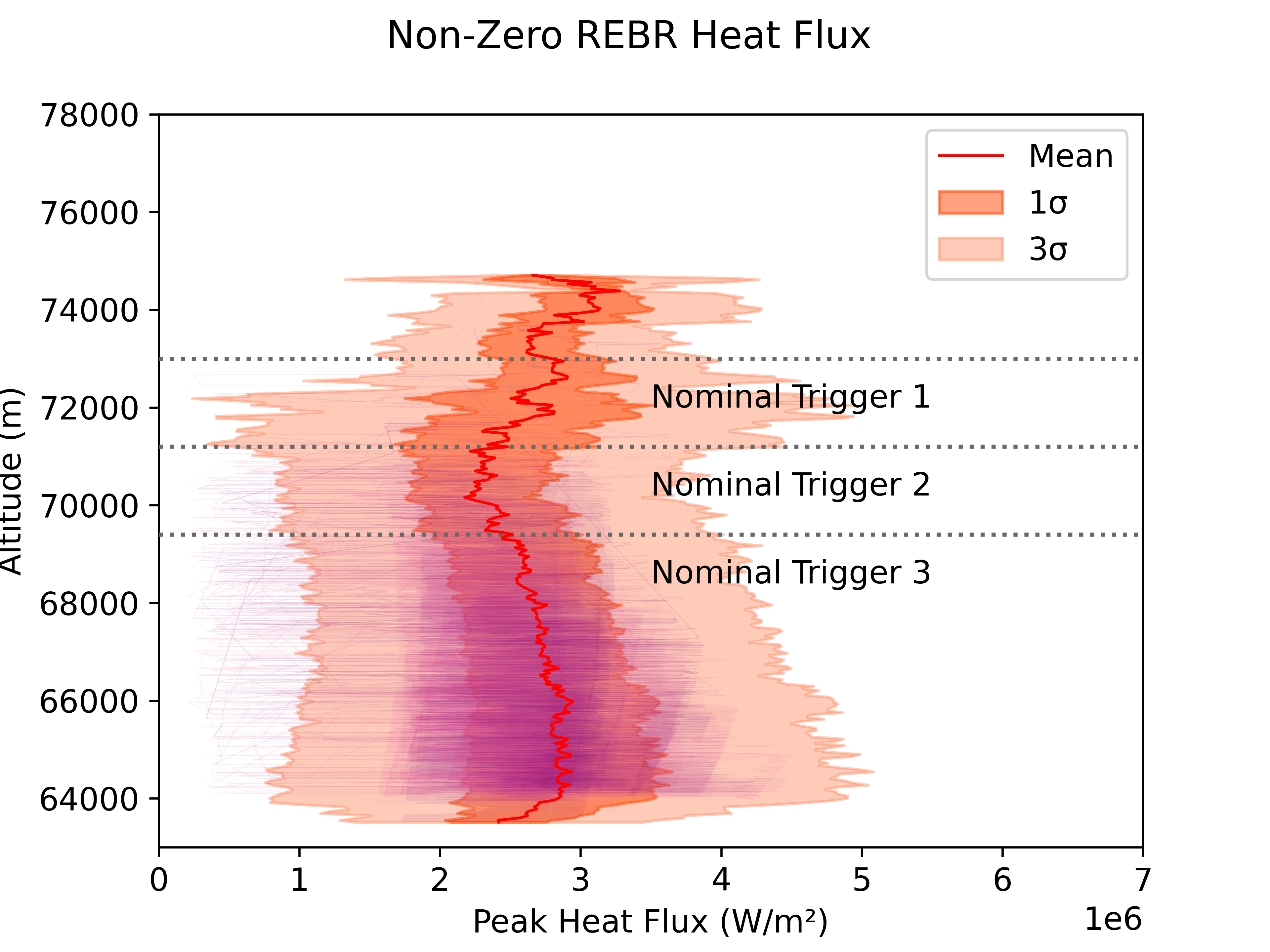}
        \caption{Peak Heat flux incident on the REBR}
        \label{fig:heatflux}
\end{figure}
\begin{figure}[h]
        \centering
        \includegraphics[width=0.75\textwidth]{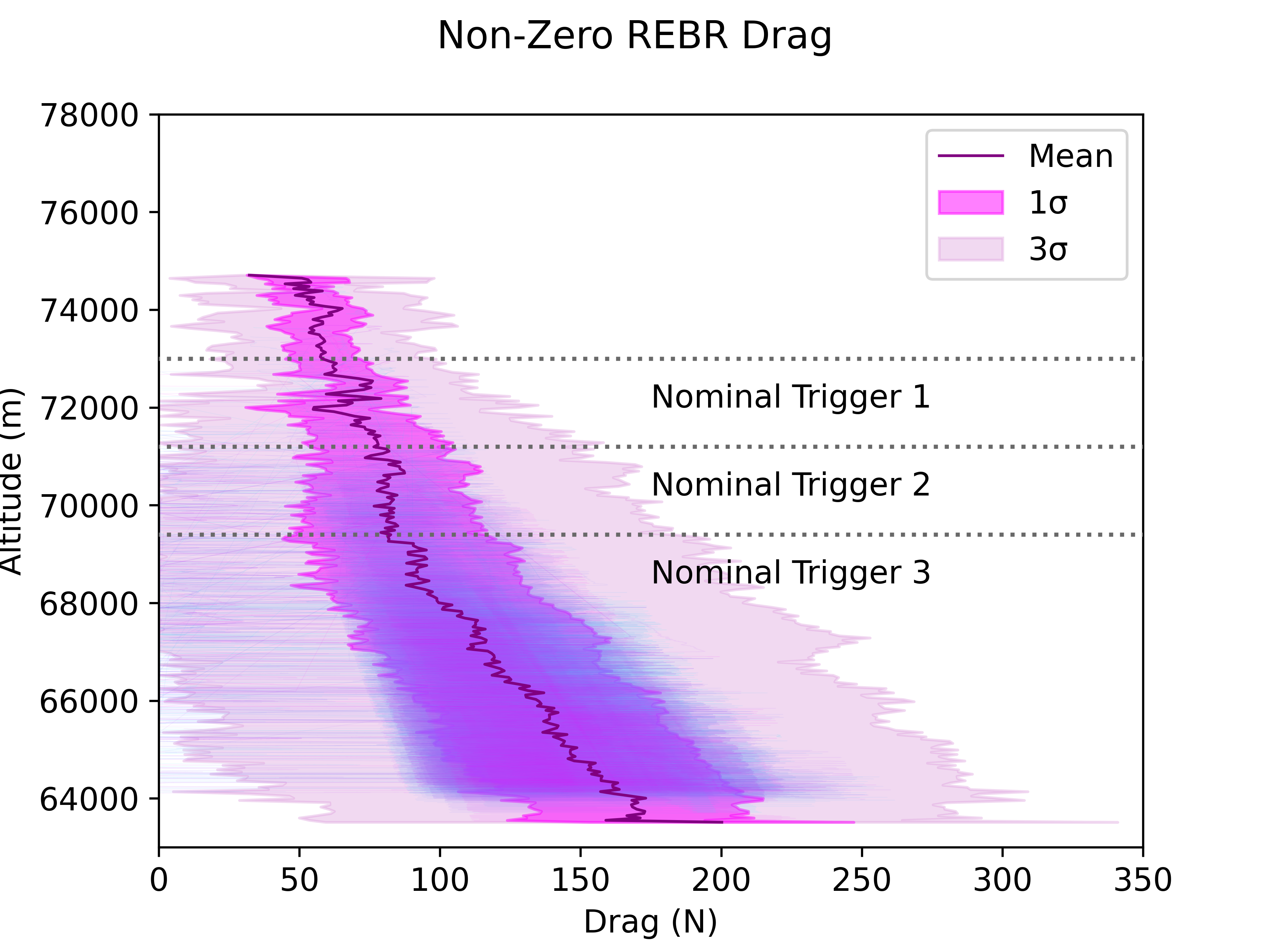}
        \caption{Drag force incident on the REBR}
        \label{fig:drag}
\end{figure}

This bimodality is evident when looking at relative linear velocity between the ATV equipped bay case and the REBR case\autoref{fig:vel_pair}.
 Comparison of the non-zero temperature increase, heat flux and drag shows expected behaviour for a re-entering body (\autoref{fig:temp}, \autoref{fig:heatflux}, \autoref{fig:drag}). For the vast majority of situations where the REBR is still within the cargo bay cavity it is shadowed from the flow but exceptions to this rule-of-thumb can be seen by observing the small number of very small heat fluxes and drag forces present in \autoref{fig:heatflux} and \autoref{fig:drag}. This again motivates interpretation of these results as representing two different dynamical schema, one dominated by multi-body dynamics with collision and contact and the other dominated by hypersonic aerodynamics.\par 
Through consideration of this data, it is the opinion of the authors that the bouncing case of early REBR detachment due to mechanical loads prior to any exposure to re-entry heating better explains the available data than alternate hypothesised event sequences. However it should be noted that, like all modelling exercises, there are limitations to the work carried out here. Alongside investigating afore-mentioned uncertainties in terms of representation of the true internal state of the cargo bay and necessarily limited representations of the complex flow physics of cavities in hypersonic flow, future work could investigate the effects of frictional contact on the case as well as increasing the fidelity of other pertinent models or including representations of additional phenomena such as potential explosive events.
\section{Conclusions}\label{sec:conc}
This work investigated the breakup dynamics of re-entering spacecraft through the lens of collision modelling and multibody dynamics as a method for exploring the nature of the behaviour of ``data recorder''-type sensors. The relevant disciplinary models were introduced, with specific focus on the methods for collision modelling in terms of binary search time of impact calculation, PGS-LCP collision resolution and split-impulse stabilisation.\par
The modelling paradigm was applied to the case of the Edoardo Amaldi ATV3 re-entry and associated REBR4 capsule according to a nominal entry profile. Through this analysis the chaotic behaviour of the REBR motivated the construction of a hypothesis on the kinds of feasible event sequences that occurred during the ATV3 re-entry, namely that the REBR was detached from its mounting and free to move in the cargo bay prior to the main breakup of the bay and was ``bouncing'' inside the cargo bay. It was additionally noted that the complex physics of cavities in the flowfield could create attractive regions that merit further study.\par
In order to evaluate this hypothesis, a Monte Carlo campaign was carried out to explore the configuration space of breakup events and account for uncertainty in initial spacecraft attitude and atmospheric density. From the results of this campaign, bimodalities were observed in the dynamics of the REBR between those instances that remained in the cargo bay cavity and those that escaped the debris cloud. By considering the data in terms of position and attitude dynamics alongside aerothermodynamic results, and by directly comparing angular velocity measurements to those recorded by the REBR it was concluded that the bouncing hypothesis better explains the available data than other hypotheses.\par
Recommendations were made that future work could investigate higher fidelity representations of the ATV's mass distribution, flow physics and contact modelling alongside accounting for potential explosions.\par
The computational model of the ATV3 and REBR used to generate the data present in this work is available from the corresponding author upon reasonable request.
\section{Acknowledgements}
The authors would like to acknowledge the financial support of the European Space Agency (ESA) through the grant Protecting Earth from the Uncertainty of Space Debris Re-Entry (PEACE) ESA Contract No. 4000143854, 2024-2027. Results were obtained for this study using the ARCHIE-WeSt High Performance Computer (https://www.archie-west.ac.uk) based at the University of Strathclyde.
\bibliography{JSR_UQ}
\appendix
\section{Detailed Material Description of ATV Model}\label{ap:materials}
\autoref{tab:ATV_material} details the materials used for components in the model of the ATV3 used in this work. It should be noted that in some cases, due to geometric considerations leading to differences in wall-thickness between the model and reality (i.e cases where volumes are fully filled as opposed to thin-walled or vice versa), that the represented component density diverges from the density of the bulk material the component is assumed to be made from. 
\begin{table}[h]
    \caption{ATV3 Model Material Properties}
    \label{tab:ATV_material}
    \footnotesize
    \centering
    \begin{tabular}{lllll}
        \toprule
        \textbf{Component}&\textbf{Material}&Bulk Material Density(kg/m3) & Component Density(kg/m3)& Mass(\unit{\kilogram})\\
        \midrule
        Russian Docking System & AA7075 & 2813.00 & 450.00 & 161 \\
        Equipped Pressurised Module(EPM) "Cap"& AA7075 & 2813.00 &  200.00 & 472 \\
        EPM Walls & Aluminium Alloy & 2700.00 & 750.00 & 1214 \\
        EPM Racks & Aluminium Alloy & 2700.00 & 297.15 & 1387 $\times 4$\\
        EPM Rack Mounts & AA7075 & 2813.00 & 2813.00 & 299 $\times 4$\\
        \midrule
        Equipped External Bay(EEB) Walls & Aluminium Alloy & 2700.00 & 750.00 & 573 \\
        EEB Baseplate & AA7075 & 2813.00 & 2813.00 & 687 \\
        EEB Yellow Tanks & TiAl6V4 & 4417.00 & 4417.00 & 13 $\times 6$ \\
        EEB Green Tanks & AA2219 & 3000.00  & 3000.00 & 50 $\times 4$ \\
        EEB Small Black Tanks & CFRP & 1580 & 358.23 & 4 $\times 3$ \\
        EEB Large Black Tanks & CFRP & 1580 & 13425.68 & 286 $\times 3$ \\
        \midrule
        Equipped Avionics Bay &  Aluminium Alloy & 2700.00 & 750.0 & 418 \\
        \midrule
        Equipped Propulsion Bay (EPB) Walls & Aluminium Alloy & 2700.00 & 750.0 & 620 \\
        EPB Upper Tank Plate & AA7075 & 2813.00 & 2813.00 & 543 \\
        EPB Lower Tank Plate & AA7075 & 2813.00 & 2813.00 & 1020 \\
        EPB MMH Tanks &  TiAl6V4 & 4417.00 & 4544.71 &  309 $\times 2$ \\
        EPB MON Tanks &  TiAl6V4 & 4417.00 & 4794.80 &  326 $\times 2$ \\
        EPB COPV Tanks & COPV & 1580 & 4221.13 & 110.00 $\times 2$ \\
        \midrule
        Separation and Distancing Module(SDM) Walls& Aluminium Alloy & 2700.00 & 750.00 & 196 \\
        SDM Thruster Plate & AA7075 & 2813.00 & 2813.00 & 396 \\
        Thruster Assembly & Inconel & 8190.00 & 8190.00 & 45 $\times 4$ \\
        Thruster Nozzle & Inconel & 8190.00 & 8190.00 & 78 $\times 4$ \\
        \midrule
        Solar Array Panels & Panel Material & 50.0 & 50.0 & 8 $\times 16$ \\
        Solar Array Joints & AA7075 & 2813.00 & 2000.00 & 5 $\times 12$ \\
        Solar Array Mounts & AA7075 & 2813.00 & 2000.00 & 9 $\times 4$ \\
        \midrule
        REBR & Carbon Ablator & 1580 & 562.00 & 4.00 \\
        \bottomrule
  \end{tabular}
  \end{table}
\end{document}